\documentclass[a4paper]{iopart}
\usepackage[round]{natbib}
\usepackage{subcaption}
\usepackage{lscape}
\expandafter\let\csname equation*\endcsname\relax
\expandafter\let\csname endequation*\endcsname\relax
\usepackage{amsmath}
\usepackage{graphicx}
\usepackage{threeparttable}
\usepackage{gensymb}
\usepackage{dblfloatfix}
\usepackage{diagbox}
\usepackage{booktabs}
\usepackage{lscape}
\usepackage{lineno}

\usepackage{setspace} 

\begin{document}

\title{Relating EEG to continuous speech using deep neural networks: a review.}
\author{Corentin Puffay$^{1,2, *}$, Bernd Accou$^{1,2}$, Lies Bollens$^{1,2}$,  Mohammad Jalilpour Monesi$^{1,2}$, Jonas Vanthornhout$^{1}$,  Hugo Van hamme$^2$, Tom Francart$^{1, *}$}
\address{
  $^1$KU Leuven, Dept. Neurosciences, ExpORL, Leuven, Belgium\\
  $^2$KU Leuven, Dept. of Electrical engineering (ESAT), PSI, Leuven, Belgium\\
  $^*$Authors to whom any correspondence should be addressed.}
\ead{corentin.puffay@kuleuven.be, bernd.accou@kuleuven.be, lies.bollens@kuleuven.be, mohammad.jalilpourmonesi@kuleuven.be, jonas.vanthornhout@kuleuven.be,  hugo.vanhamme@kuleuven.be, tom.francart@kuleuven.be}

\begin{abstract}

\textit{Objective.} When a person listens to continuous speech, a corresponding response is elicited in the brain and can be recorded using electroencephalography (EEG). Linear models are presently used to relate the EEG recording to the corresponding speech signal. The ability of linear models to find a mapping between these two signals is used as a measure of neural tracking of speech. Such models are limited as they assume linearity in the EEG-speech relationship, which omits the nonlinear dynamics of the brain. As an alternative, deep learning models have recently been used to relate EEG to continuous speech.\\
\textit{Approach.}
This paper reviews and comments on deep-learning-based studies that relate EEG to continuous speech in single- or multiple-speakers paradigms. We point out recurrent methodological pitfalls and the need for a standard benchmark of model analysis.
\textit{Main results.} We gathered 29 studies. The main methodological issues we found are biased cross-validations, data leakage leading to over-fitted models, or disproportionate data size compared to the model's complexity. In addition, we address requirements for a standard benchmark model analysis, such as public datasets, common evaluation metrics, and good practices for the match-mismatch task.
\textit{Significance.} We present a review paper summarizing the main deep-learning-based studies that relate EEG to speech while addressing methodological pitfalls and important considerations for this newly expanding field. Our study is particularly relevant given the growing application of deep learning in EEG-speech decoding.\\

\end{abstract}

\section{Introduction}

Electroencephalography (EEG) is a non-invasive method that measures the electrical activity in the brain. When a person listens to speech, the EEG signal measured has been shown to contain information related to different features of the presented continuous speech. We can relate these speech features to the EEG activity using machine learning models to investigate if and how the brain processes continuous speech.
Reasons for doing this include (1) understanding neural mechanisms of speech processing in the brain; (2) objectively measuring processes in the brain related to speech processing, which in turn are useful for research and clinical diagnostics of hearing; and (3) designing auditory prostheses that incorporate attention decoding and use it to steer noise suppression to improve speech understanding in so-called cocktail party scenarios.\\ 

The current literature presents two approaches, either considering a single, or multiple sound sources when fitting a stimulus-response model.\\
The single sound source approach usually aims to quantify the time-locking of a brain response to a single speech source, often referred to as neural tracking. Neural tracking of speech can be used in multiple applications, notably to model the speech processing in the brain but also as an objective measure of hearing or understanding speech.\\
To objectively measure neural tracking, two main tasks have been used: the match-mismatch task, and direct regression of stimulus or EEG\\
In the match-mismatch (MM) paradigm \citep{DECHEVEIGNE2018} a model is trained to associate a segment of EEG with the corresponding segment of speech, and the accuracy obtained on this task is defined as a measure of neural tracking. Variations of the MM task were implemented with either one \citep[e.g.,]{DECHEVEIGNE2018, Monesi2020} or more \citep[e.g.,]{Puffay2022, Accou2021PredictingSI, Monesi2021INTERSPEECH} speech segment candidates to associate the EEG segment with.\\
EEG can also be related to speech in a reconstruction/prediction (R/P) task, or direct regression. In this case a stimulus feature is reconstructed from the EEG (or the EEG predicted from the speech, respectively), and correlated to the original signal. This relates to the commonly used linear backward (or forward, respectively) models.\\
In another linear approach, canonical component analysis (CCA), the stimulus and EEG are projected to separate subspaces and correlated in those subspaces  \cite{DECHEVEIGNE2018}.\\

When more than one speaker talks simultaneously, the brain of a listener must cope with multiple speech sources. One of the main challenges arising from this scenario, so called auditory attention decoding (AAD), is to detect which speech source was attended by the listener. The interest in this topic is two-fold: it provides a basis to overcome current hearing aid limitations in cocktail party scenarios, but also to investigate attention mechanisms in the brain.
In a typical AAD paradigm, multiple speech streams are presented to the listener who is required to attend to one speaker. A typical recording comprises several trials separated by short breaks, in which the subject is instructed to maintain focus on a designated speaker throughout each trial. Multiple trials are recorded during an experiment in order to switch attention between speakers, sometimes also with changing acoustic conditions.
To measure attention decoding accuracy, two main tasks are used: speaker identity and directional focus classifications.
In the speaker identity (SI) classification task, a model is trained to decode which speech stream is attended given the EEG response and N given speech streams.
Another approach is to decode the directional focus of attention (DFA). It presents advantages such as avoiding separating speech sources, but also enabling models to use brain lateralization (i.e. the directional focus is encoded spatially in the brain), which is an instantaneous spatial feature, rather than a temporal feature.
Classification accuracy measures auditory attention for both tasks.\\

Although correlation is often used as the basis for decision in the MM or AAD tasks, it also plays the additional role of a metric of quality for the R/P or CCA tasks. However, comparison of correlations across experiments is delicate if the preprocessing of the target signals is different. As an example, predicting an EEG in a narrow frequency band is easier than predicting the whole EEG spectrum. This implies that to maximize correlation, one can decide to filter the EEG in a given frequency band, while some speech-related information is contained in the filtered-out EEG (\cite{DeCheveigne2021}). We provide a detailed criticism of correlation as a measure of quality of a model in Section \ref{correlation_score}.\\
MM and AAD tasks do not suffer from the issues that plague correlation in that they are objective tasks \citep{DECHEVEIGNE2018}. The objective tasks' performance is impaired by the loss of information due to preprocessing steps such as filtering of the input data or EEG channel re-referencing, which provides them more realistic modeling abilities.\\

Linear models are limited in that they assume linearity between the EEG and speech signals, inadequately fitting the nonlinear nature of the auditory system. For example, it is well known that depending on the level of attention and state of arousal of a person response latencies can change \citep{Ding2012}, which cannot be modeled with a single linear model.\\
Deep neural networks (DNNs) have been recently introduced to this field. Many studies have shown the ability of deep learning models to relate EEG to speech (see Section \ref{Review}), be it for neural tracking assessment \citep[e.g.,][]{Katthi2021DeepMC, Accou2021ModelingTR, Monesi2020, Thornton_2022} or to decode auditory attention \citep[e.g.,][]{deTaillez2020MachineSpeech, Ciccarelli2019, kuruvila2021extracting}.
On certain tasks, DNNs have outperformed linear decoder baselines \citep[e.g.,][]{Accou2023, Accou2021ModelingTR, Monesi2020, Puffay2023.03.30.534911}, but it is still not a general finding.\\

In this paper, we summarize the methods present in the literature to relate EEG to continuous speech using deep learning models. 
In Section \ref{Review}, we first review different experiment steps of the gathered studies and the different approaches chosen by authors. These steps include the task used to relate EEG to speech, the network architecture used, the dataset's nature, the preprocessing methods employed, the dataset segmentation, and the evaluation metric. 
In Section \ref{Pitfalls}, we then address the methodological pitfalls to avoid when using such models and we recommend to establish a standard benchmark for models' analyses.
In \ref{multiple_source_appendix} and \ref{single_source_appendix}, we summarize the results of each study individually for multiple sound source and single sound source respectively.

\section{Review of deep-learning-based studies to relate EEG to continuous speech}\label{Review}

\begin{table*}[htpb!]
 \caption{Search queries for each search engine during paper collection.}
\hspace*{-1cm}\begin{tabular}{@{}ll@{}}\toprule

  Search engine  & Search query \\ \midrule 
  Google Scholar &  ("EEG" OR  "Electroencephalography" OR "Electroencephalogram") AND speech \\
  & AND ("deep learning" OR "deep neural networks")\\ 
  
 \rule{0pt}{4ex} IEEE Xplore & (("All Metadata":EEG) OR ("AlklMetadata":Electroencephalogram))\\
 & AND ("All Metadata":speech)\\
 & AND (("All Metadata":deep neural networks)\\
& OR ("All Metadata": non-linear) OR ("All Metadata": nonlinear) )  \\

\rule{0pt}{4ex} Science Direct & (EEG OR Electroencephalogram)  AND ("continuous speech" OR "natural speech")\\
& AND ( "neural network")) \\

\rule{0pt}{4ex} Pubmed & (EEG OR Electroencephalography OR Electroencephalogram)\\
& AND (speech)  AND ( deeplearning OR deep learning OR neural networks)\\
&  NOT (imagined[title]) NOT  (motor[title]) NOT emotion[title] \\

 \rule{0pt}{4ex}  Web of Science & ((((((((TS=(("EEG" or "Electroencephalography" or "Electroencephalogram")))\\
 & AND TS=(("speech" or "audio" or "auditory" )))\\
 & 
   AND TS=(("artificial neural network*" or "ANN" or "deep learning" or\\
   & "deeplearning" or "CNN" or "convolutional" or  "recurrent"\\
   & or "LSTM"))) NOT TI=("imagined" or "motor imagery" or "parkinson"))\\
   & NOT TI=("emotion")) NOT TS=(("dysphasia" or "alzeimer*")))\\
   & NOT TS=("seizure"))) AND  DOP=(2010-01-01/2022-05-15)\\

\\\bottomrule
\end{tabular}
\vspace{0.3cm}
 \label{tab:search_query}
\end{table*}

Using Google Scholar, IEEE Xplore, Science Direct, Pubmed and Web of Science, we collected papers using search queries reported in Table \ref{tab:search_query}. As a last step, we pruned the  selection manually to exclude studies not including EEG data, continuous speech stimuli or deep learning models, and stopped searching for new papers in December 2022. 

In this section, we go through different features of the gathered studies including the task used to relate EEG to speech, the different architectures used, the dataset's nature, the preprocessing methods employed, the dataset segmentation, and the evaluation metrics. More detailed summaries of individual studies can be found in \ref{multiple_source_appendix} and \ref{single_source_appendix}.

\subsection{Tasks relating EEG to speech}\label{sec:task}

To relate EEG to speech, we identified two main tasks, either involving a single speech source or multiple simultaneous speech sources.\\

In the gathered papers including the single sound source approach, we identified two main tasks: the MM and the R/P tasks (see Table \ref{table2}).
We report four studies in which the MM task was used \citep{Monesi2020, Monesi2021INTERSPEECH, Accou2021ModelingTR, Accou2021PredictingSI}. On the other hand, five studies used an R/P task, including four with backward modeling \citep{KrishnaEUSIPCO2020, Krishna2021NER, Sakthi2019, Thornton_2022}, and one with forward modeling \citep{KrishnaEUSIPCO2020}. Variations of CCA have also been explored in two studies \citep{Katthi2021DCA, Katthi2021DeepMC}.\\
The MM and the R/P tasks were the most used methods, however some studies used different tasks to relate EEG to speech such as semantic incongruities classification \citep{Motomura2020},  or sentence classification \citep{Sakthi2021}. \cite{Bollens2021} utilized solely an embedded representation to classify segments of speech. \\

In the gathered papers including the multiple sound sources approach, we identified two main tasks: the SI and the DFA tasks (see Table \ref{table1}).\\
We report a majority of studies implementing the SI task.
From 2016 to 2020, only three studies were reported \citep{shree2016novel, deTaillez2020MachineSpeech, tian2020auditory}, however since 2021 this has increased a lot \citep{su2021auditory, kuruvila2021extracting, lu2021auditory, zakeri2021supervised, Vandecappelle2021, Hosseini2021ICASSP, xu2022auditory, xu2022decoding, Thornton_2022}.\\
On the other hand, we only found one study that uses deep learning to decode the locus of attention \citep{Vandecappelle2021}, it could thus be worthwhile to consider this as a potential avenue for future AAD research.\\

\subsection{Model architectures}\label{sec:architecture}

The field of deep learning is evolving rapidly, and constantly providing novel architectures. Multiple layer types were integrated to AAD and single speech source decoding models. Globally, architectures to solve the tasks mentioned in Section \ref{sec:task} were inspired from other fields (e.g., automatic speech recognition, ASR). We provide a more in-depth description of each architecture in \ref{appendix_architecture}.\\

Early attempts used general regression neural networks (GRNNs) \citep{shree2016novel} or fully-connected neural network (FCNN) models \citep{deTaillez2020MachineSpeech}. As fully connected layers involve a high complexity, possibly leading to overfitting and high computation costs, later studies implemented convolutional neural network (CNN)-based models \citep{Ciccarelli2019, tian2020auditory, Thornton_2022, Vandecappelle2021}.\\
As an attempt to allow context to be used in a non-linear and/or non-stationary fashion,  models with recurrent layers such as long-short term memory (LSTM) \citep{Monesi2020,Monesi2021INTERSPEECH, kuruvila2021extracting, lu2021auditory, xu2022auditory}, Bi-LSTM \citep{zakeri2021supervised}, gated recurrent unit (GRU) \citep{Krishna2020, Krishna2021NER, KrishnaEUSIPCO2020, Sakthi2021} or Bi-GRU \citep{Motomura2020} were implemented.\\
To enable the model to provide more weight to certain time points \citep{Motomura2020, KrishnaEUSIPCO2020}, or certain EEG electrodes \citep{su2021auditory}, channel attention mechanisms were integrated into the models. In one study \citep{xu2022decoding},  channel attention was integrated into a transformer, a well-known architecture firstly introduced in natural language processing tasks that allows parallel computation (to reduce training time), and reduces performance drops due to long dependencies \citep{AttentionisAllyouneed}.\\
Other popular model types such as generative adversarial networks (GANs) \citep{Krishna2021NER}, or autoencoders (AEs) \citep{Bollens2021, Hosseini2021ICASSP} were utilized.  AEs find a compressed meaningful representation of a signal. They can be constrained to extract speech-related information in EEG, hence working as a denoiser.

\subsection{Datasets}

As explained in Section \ref{sec:architecture}, deep learning architectures have very useful properties to relate EEG to speech. Compared to linear models, they often have a high number of parameters, which means lots of data are required to train properly, and to avoid overfitting.
As collecting EEG data is tedious and time-consuming, research groups often work with their own small datasets, sometimes containing a few minutes of speech or less per subject \citep{lu2021auditory, shree2016novel, Krishna2020, KrishnaEUSIPCO2020, Krishna2021NER, tian2020auditory}. The other studies we reported used at least 30~min of data per subject, while some even published their datasets \citep{das_neetha_2019_3377911, fuglsang_soren_a_2018_1199011, K3VSND2023}. We discuss the importance of the dataset and generalization in Section \ref{public_dataset}.

\subsection{Preprocessing}

Once a dataset is available, both the EEG and the presented speech can be preprocessed in various manners.

\subsubsection{EEG}

For EEG preprocessing, most of the studies we reviewed start with filtering the signal, first with a high-pass filter to remove any unwanted DC shifts or slow drift potentials, secondly with a low-pass filter to remove high frequencies as the SNR becomes lower in higher frequency ranges. In linear studies, models typically use low frequencies (e.g., between 0.5 and 8~Hz), while some deep learning studies report benefits from including higher frequencies \citep{Puffay2022}.
A re-referencing step can then be added, typically by subtracting the mean over all channels from each individual channel. This contributes to increase the signal-to-noise ration (SNR).
Downsampling  is commonly performed to reduce the computational time during training. It is often done in accordance with previous filtering to avoid temporal aliasing as mentioned in \cite{Crosse_review}. Typical sampling rates are 128 or 64~Hz.
Finally, an artifact removal algorithm based on different methods such as multi-channel Wiener filtering (MWF) \citep{Somers2018}, or independent component analysis (ICA) \citep{HYVARINEN2000411} is used to remove different artifacts (e.g., eye-blink, neck movement).\\ 

The majority of the studies we reviewed provided their models with EEG signals preprocessed as stated above. However, two of them engineered more specific features from EEG, such as a latent representation optimized through the training of an AE \citep{Bollens2021}, or source-spatial feature images \citep{tian2020auditory}.\\

While it is not specific to deep learning, many trained models can subsume steps that would traditionally be considered as preprocessing, which makes it challenging to quantify the impact of preprocessing on a given model's performance. For more details, we invite readers to examine individual studies. Extensive considerations about preprocessing are reported by \cite{Crosse_review}.\\

\subsubsection{Speech}

From the raw speech signal, it is common to investigate the neural tracking of different features of speech. Most studies we report here used acoustic features, such as the temporal envelope \citep[e.g,][]{Ciccarelli2019, su2021auditory, deTaillez2020MachineSpeech, lu2021auditory, xu2022auditory, xu2022decoding}, or the Mel spectrogram \citep[e.g.,][]{Krishna2020, KrishnaEUSIPCO2020, Krishna2021NER, Monesi2021INTERSPEECH, kuruvila2021extracting}. A study even used the fundamental frequency of the voice -f0- \citep{Puffay2022}.\\
To investigate the processing of different speech units in the brain, higher-level speech features at the level of the sentence \citep{Motomura2020, Sakthi2021}, word \citep{Monesi2021INTERSPEECH} or phonemes \citep{Sakthi2021, Monesi2021INTERSPEECH} were also used.\\

As mentioned by \cite{Crosse_review}, these preprocessing steps should be conducted on the entire dataset. High-pass filtering on minute-long trials can introduce substantial edge artifacts. On the other hand, if a hold-out set is selected for model evaluation, a normalization prior to segmentation can leak information from the hold-out set to the training set, which will bias the evaluation performance.

\subsection{Data segmentation}

The training paradigm is crucial and should be carefully performed to avoid biases and overfitting. Typically, three separate sets are employed: the training set, the validation set, and the test set. The training set is utilized to train the model by adjusting the weights and biases to minimize the loss function. The validation set is employed to tune the hyperparameters during the training process. Lastly, the test set, which remains unseen during training, is utilized to evaluate the final performance of the model without any bias.\\
In some gathered studies, cross-validation was employed \citep[e.g.,][]{Ciccarelli2019}. Cross-validation is a method that iterates over different data segmentations to train and validate a given model. In a more simplistic manner, a single cross-validation iteration was also occasionally performed \citep[e.g.,][]{Monesi2021INTERSPEECH, deTaillez2020MachineSpeech}.\\

In AAD tasks, trials are defined as a period of time during which a subject attends to a target speaker, and are labelled accordingly (e.g., left/right). The split can therefore be done in different manners, notably within \citep[e.g.,][]{lu2021auditory} and between trials \citep[e.g.,][]{Thornton_2022}. The split choice can make the evaluation sensitive to overfitting, as discussed in Section \ref{AAD}.

\subsection{Evaluation metric}

For single sound source paradigms, various metrics are employed. Classification metrics are used, such as the match-mismatch accuracy \citep{Monesi2020, Monesi2021INTERSPEECH, Accou2021ModelingTR, Accou2021PredictingSI}, subject-classification accuracy \citep{Bollens2021} or sentence classification accuracy \citep{Motomura2020}. 
For R/P studies, the main metric we noted is Pearson correlation \citep{Katthi2020, Katthi2021DeepMC, Thornton_2022}, while one research group used root-mean-squared error (RMSE) and Mel cepstral distortion (MCD) \citep{Krishna2020, KrishnaEUSIPCO2020, Krishna2021NER}.\\ 

While in a multi talker paradigm various intermediate metrics can be used to select the attended speaker or locus, in all studies we report the attention decoding (i.e., speaker identity or direction classification) accuracy is used. The only metric differences are the trial length and the chance level defined by the number of sound sources.\\

The use of multiple metrics is problematic when comparing model performances across studies. For instance, even when the same metric is used (e.g., Pearson correlation), it does not guarantee a fair comparison (see Section \ref{correlation_score}). The MM and AAD tasks both yield objective percent-correct scores and therefore do not suffer from the null-distribution variations obtained with different parameters.

\begin{landscape}
\begin{table*}
\hspace{-4cm}\begin{threeparttable}
\vspace{2.5cm}\begin{tabular}{@{}llllllll@{}}\toprule

   Article & Architecture & Feature & Task & Split (train/test/val) & Window & Performance & subjects - stimuli \\ \midrule 

    \cite{Sakthi2019} & LSTM & E/S & R, O & 80/20\% (sub) & x  & 0.20-0.22/0.09-0.12 & $15 - 30\times1~min$\\  

     \cite{Krishna2020} & GRU & MFCC & R & 80/10/10\% (sub) & x & MSE, MCD  & $4 - 70\times2~words$\\

 \cite{Monesi2020} & LSTM & E & MM & 80/10/10 (stim) & 5s; 10s  & 80; 85\% & $90 - 10\times~12~min$  \\ 

  \cite{Motomura2020} & bi-GRU+Att & Se & O & 13/2/4 (sub) & x  & 63.5\%  & $19 - 200~sentences$\\

    \cite{KrishnaEUSIPCO2020} & GRU, GAN & MFCC & P & 80/10/10\% (sub) & x  & MSE, MCD  & $4 - 70\times2~words$\\
 
    \cite{Monesi2021INTERSPEECH} & LSTM & M & MM & 80/10/10 (stim) & 5s & 84\% & $90 - 10\times~12~min$ \\ 
    
    \cite{Accou2021ModelingTR} & Dilated CNN & E & MM & 80/10/10 (stim) & 10s & 90.6\% & $48 - 10\times~12~min$ \\
    
    \cite{Accou2021PredictingSI} & Dilated CNN & E & MM & 80/10/10 (stim) & 10s & 90.6\% & $48 - 10\times~12~min$ \\
    
    \cite{Katthi2021DeepMC} & FCNN & E & R/P & CV (stim) & x & from 0.2 to 0.4 & $6 - 20\times~160~s$ \\
    
     \cite{Katthi2021DCA} & AE+FCNN & E & R & CV (stim) & x & 0.27 to 0.344 & $8 - 20\times~160~s$ \\

    \cite{Krishna2021NER} & GRU+Att. & MFCC & R & 80/10/10\% (sub) & x  & MSE, MCD  & $4 - 70\times2~words$\\

         \cite{Sakthi2021} & LSTM/GRU & Se/P & O &  70/30\% (stim) & x & F1 & $16 - 600~sentences$\\      
    
     \cite{Bollens2021} & AE & None & O & 80/10/10 (stim) & 500~ms  & 98.96\% - 62.91\%  & $100 - 8\times15min$\\

        \cite{Puffay2022} & CNN & E, f0 & MM & 80/10/10 (stim) & 2~s & 65\% to 75\% & $60 - 8\times15min$\\
         
             \cite{thornton2022robust} & FCNN & E & R & 9/3/3 (sub) & x  & 0.255 to 0.344 & $13 - 40~min$\\\bottomrule
             
\end{tabular}
\vspace{0.3cm}
 \caption{Key figures for single speech source papers sorted chronologically. E=envelope, M=multiple, f0=fundamental frequency of the voice, S=spectogram, Se=sentence, P=phoneme, MM=Match-mismatch, R=Reconstruction task, P=prediction, O=Other, CV=cross-validation, stim=within stimulus, sub=within subject. The performance values reported depend on the task: for MM and D it is a classification accuracy (\%) and for R it is a Pearson correlation value. Study-specific metrics are specified in the table as names instead of numbers.}
\label{table2}
\end{threeparttable}
\end{table*}
\end{landscape}

\begin{landscape}
\begin{table*}

\hspace{-4cm}\begin{threeparttable}
\vspace{2.5cm}\begin{tabular}{@{}llllllll@{}}\toprule

   Article & Architecture & Feature & Task & Split (train/val/test) & Window (s) & Accuracy(\%)  & subjects - stimuli \\ \midrule 

\cite{shree2016novel} & GRNN & E & SI & 50/0/50 (unclear) & 60 & 99.05 (locus)   & $20 sub - 3~min$ \\

 \cite{Ciccarelli2019} & CNN & E & SI & 60/10/10 (unclear) & 10 & 81  & $11 - 40~min$   \\ 

    \cite{deTaillez2020MachineSpeech} & FCNN  & E & SI & 80/10/10  (dataset)  & 60-10-5 &  97.6-86-79  & $16 - 50~min$ \\

     \cite{tian2020auditory} & CNN & EEG & SI & x & x & x  & $42 sub - 120\times2~s$ \\

    \cite{su2021auditory} & CA + CNN & E & SI & 60/20/20 (subject) & 0.1 to 2& 77.2 to 88.3  & $Das2019$\\

    \cite{zakeri2021supervised} & Bi-LSTM & E & SI & 63/11/26 (per trial) & 1 to 40 & 66-84  &  $12 sub - 43min$ \\

    \cite{Vandecappelle2021} & CNN & E & DFD & 3/1 (stimulus) & 1–2 & 81 (locus)   & $Das2019$ \\
    
    \cite{kuruvila2021extracting} & CNN-LSTM  & S & SI & 75/12.5/12.5 (per trial) & 2 to 5 &72-75   & $27 -30 min$, $Das2019$, $DTU$  \\

    \cite{lu2021auditory} & LSTM+FC & E & SI & 60/0/40 (per trial)  & 0.25 to 4& 96  & $21sub - 100~s$ \\

    \cite{Hosseini2021ICASSP} & AE & E & SI & 23/2/5 (trials) & x & x & $34~sub - 30~min$ \\

    \cite{xu2022decoding} & transformer & E & SI & 15/5/80 (per trial) & 0.15 & 74  & $21 sub - 40~min$ \\
    
    \cite{xu2022auditory} & LSTM & E & SI & 15/5/80 (per trial)  & 0.15& 73.35  & $21 sub - 40~min$ \\

    \cite{thornton2022robust} & CNN & E & SI & 9/3/3 (trial) & 10 & 80  & $18 sub - 10~min$ \\\bottomrule
\end{tabular}

\vspace{0.3cm}
\caption{Overview of multiple speech source papers sorted chronologically. Architecture=main layers used in the neural network, CA=channel attention; Feature=speech feature used in the model, E=envelope, S=spectrogram; Split (train/val/test)=how data were split for training, validation and test; Window (s)=decision window length in seconds; Accuracy=attention decoding accuracy in \%, subjects - stimuli= Number of subjects and length of the presented stimulus; $Das2019$=public dataset, 16 subjects, 48~min of stimuli each.  \citep{das_neetha_2019_3377911}; $DTU$=public dataset, 29 subjects, 2~h of stimuli each \citep{fuglsang2017}; SI=speaker identification; DFD=directional focus decoding. }
 \label{table1}
 \end{threeparttable}
\end{table*}

\end{landscape}

\section{Overfitting, interpretation of results, recommendations} \label{Pitfalls}

\subsection{Preamble}

In our own practice with auditory EEG we noticed how easily the deep learning models overfit to specific trials, subjects or datasets. This is mainly due to the relatively small amount of data typically available, compared to other domains such as image or speech recognition. A very careful selection of the test set is therefore needed, and the results of a number of the studies reviewed above may be overly optimistic.\\ 
In the following experiments, we demonstrate how this can happen and propose a number of good practices to avoid overfitting and how to calculate results on a sufficiently independent test set. 
As explained in the Introduction section, the nature of the data is different for single and multiple speech sources approaches, requiring distinct experiment designs. We therefore introduce two different public datasets which we use for respectively single speech source or multiple speech source experiments below.

\subsubsection{Single speech source (N=1) dataset}
For the single speech source experiments, namely subsections \ref{correlation_score}, \ref{SI_model} and \ref{MM_negative_sample} below, we select a publicly available dataset \citep{K3VSND2023}.
We selected 48 subjects: 26 subjects from a dataset that now consists of EEG data from 85 normal hearing subjects while they listened to 10 unique stories of roughly 14 minutes each, narrated in Flemish (Dutch) \citep{das_neetha_2019_3377911}, and 22 subjects who did not provide consent to publish their EEG data online.\\

\noindent As we know what performance to expect on this dataset, we use the LSTM-based model proposed by \cite{Monesi2020} trained on the MM task defined in the same study (i.e., the model must choose among two speech segments to match with the EEG segment). An exception is made for subsection \ref{correlation_score} which uses linear decoders as correlation analyses require a regression task rather than a match-mismatch task. We use linear decoders in this subsection as they face the same issue as DNNs but are computationally less expensive.

\subsubsection{Multiple speech sources (N$>$1) dataset}\label{sec:MS_dataset}
For the multiple-speech source experiment, namely subsection \ref{AAD} below, we use the publicly available dataset from \cite{das_neetha_2019_3377911}. It contains data from 16 subjects. In total, there are 4 Dutch stimuli (i.e., stories), spoken by male speakers, of 12 minutes each. Each stimulus is split up into 2 parts of 6 minutes and all stimuli are played twice, alternating the attended story. Each subject listens to 8 trials of 6 minutes.\\
In addition, the subjects listen to the first 2 minutes of each of the 4 stories three times (total: 24 minutes), which we will use as an extra held out set.\\

\begin{figure}[htpb!]
    \centering
    \includegraphics[width=\textwidth]{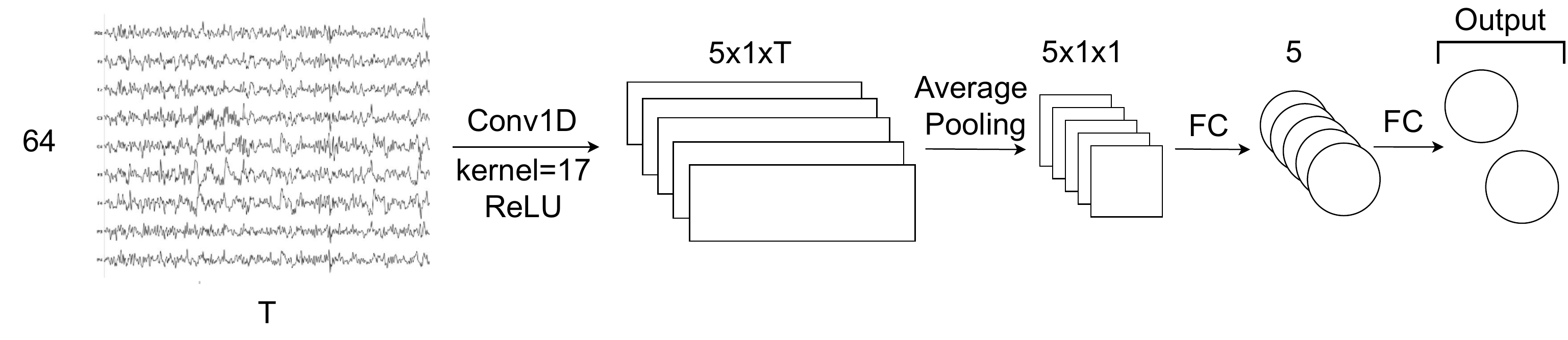}
    \caption{\textbf{Architecture of the CNN model \citep{Vandecappelle2021}.} The input has dimensions $64\times T$, representing the number of EEG channels and the segment length. Five $64\times 17$ spatio-temporal filters are shifted over the input matrix. }
    \label{fig:cnn_vandecapelle}
\end{figure}

\noindent As we know what performance to expect on this dataset, we use the CNN model proposed by \cite{Vandecappelle2021} to conduct our experiments. The architecture of this model is depicted in Figure \ref{fig:cnn_vandecapelle}.

\subsection{Selection of training, validation and test sets}\label{AAD}

When training neural networks, the split of the dataset into a training, validation, and test partition is an important aspect.\\

In a two-competing speaker scenario, the task of the model is usually to predict which one of the two speakers is the attended one and which one is the unattended one. When recording the EEG, the measurement is usually spread out over multiple trials. In each trial, the subjects have to pay attention to one of the two speakers. Then, in the next trial, they have to pay attention to another speaker, to generate a balanced dataset. Translating this into output labels, means that there is usually one label per trial, e.g., \textit{left speaker} or \textit{right speaker}. \\

A common way to split these datasets up into training/validation and testing sets is to split each individual trial into a training, validation and test segment, which are then aggregated across all trials to form the complete training, validation and test set \citep{zakeri2021supervised, lu2021auditory, xu2022auditory, xu2022decoding, su2021auditory, shree2016novel, Ciccarelli2019, deTaillez2020MachineSpeech}. With only one label per trial (left/right), the model might learn to identify the trial (e.g., left) from which the segment of EEG was taken, rather than solving the auditory attention task. If the validation and test set are taken from within the same trial, they have the same correct label, and information from the training set can leak into the validation and test set. This leads to models that seemingly perform great, but are unable to generalize and do not score well on unseen trials. 

To prevent this, we propose to always use held-out trials for the test set \citep{kuruvila2021extracting, thornton2022robust, Hosseini2021ICASSP, tian2020auditory}. If the dataset contains 10 trials, 8 could be used for training, 1 for validation, and 1 for testing. Since the trial used for testing is never seen by the model before, it cannot rely on identifying which trial the EEG segment is taken from and has to learn to identify the underlying speaker information. \\

To demonstrate the necessity of between-trial splits, we conducted experiments, using the model proposed by \cite{Vandecappelle2021}, with two different cross-validation splits of the dataset and show that this leads to substantially different results.
In both experiments, we employ a 4-fold cross-validation scheme, using three folds for training and splitting the fourth fold equally into a validation and test set. However, the key distinction lays in how we divided the dataset into folds.

In the first experiment (i.e., within-trial split), each trial is divided into 4 folds. Hence, there is a part of each trial in the training, validation and the test set.

In the second experiment (i.e., between-trial split), each fold contains unique stories, ensuring that the stories seen in training do not appear in the test set. Specifically, the folds are organized as (trial1, trial5), (trial2, trial6), (trial3, trial7), and (trial4, trial8), as outlined in Table~\ref{tab:Das2019_dataset}. Hence, each trial only appears in either training, validation or test set.

\begin{figure}[htbp!]
    \centering
    \includegraphics[width=\linewidth]{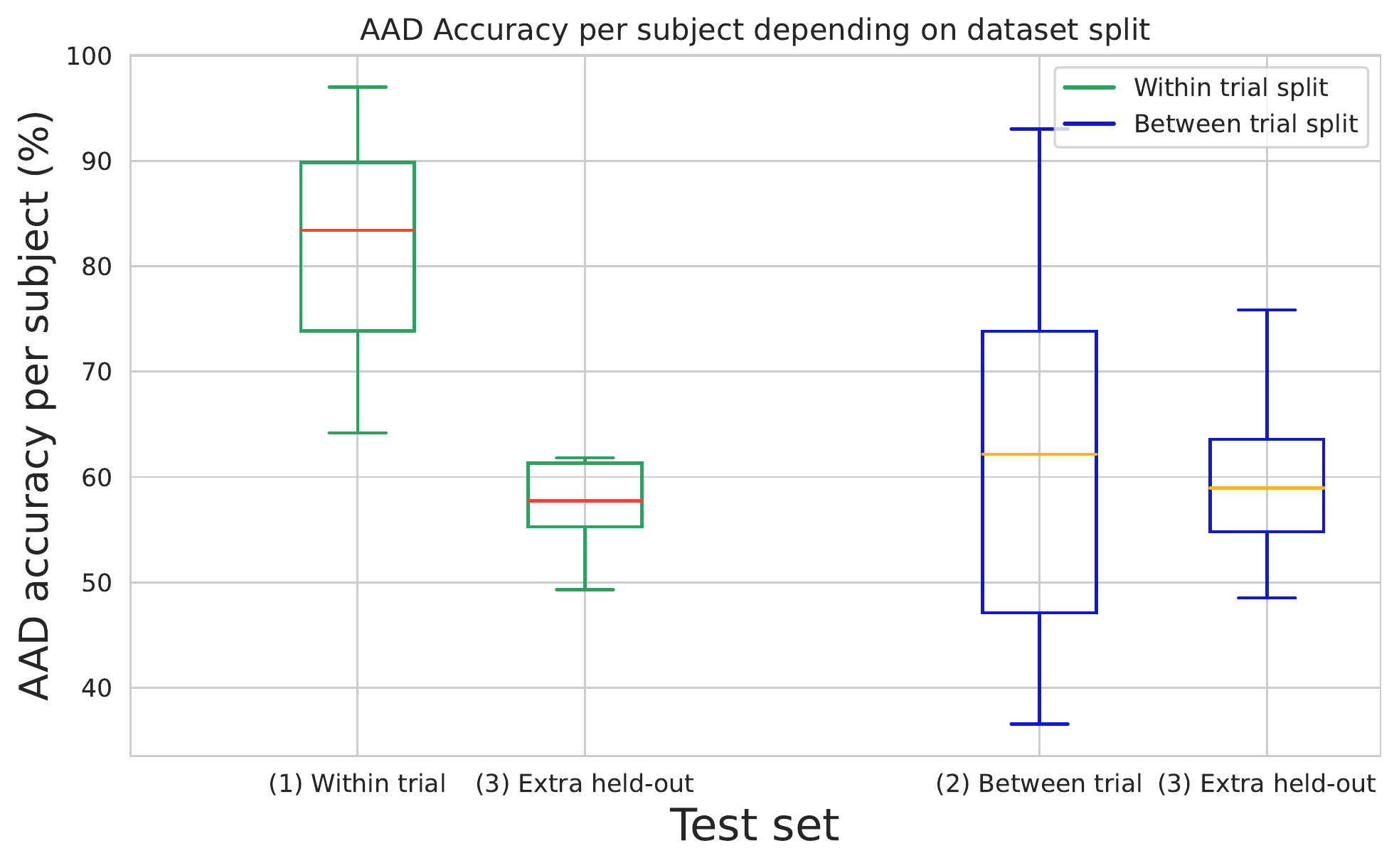}
    \caption{Results for training the model from \cite{Vandecappelle2021} using different sets of training/test. Box plots shown over 16 subjects. Each point in the boxplot corresponds to the auditory attention detection accuracy for one subject, averaged over all segments. (1) Within-trial split: for each cross-validation iteration, each trial of 6 minutes is split into 75/12.5/12.5 training/validation/test. (2) Between-trial split: for each cross-validation iteration, out of the 8 trials per subject, use 6 for training, 1 for validation and 1 for test. (3) Extra held-out set: 24 extra minutes of data for each subject, never used for training, as described in Section \ref{sec:MS_dataset}, used to compare generalization of both experiments.}
    \label{fig:aad_crossval}
\end{figure}

In addition, we test the generalizability of the models, by testing each model on the extra held-out set defined in \ref{sec:MS_dataset}, which is never used for training in any of the experiments. As explained above, this extra held out set contains three repetitions of the first two minutes of all stories, for a total of 24 minutes per subject.\\
The results of both experiments are reported in Figure~\ref{fig:aad_crossval}. The average accuracy of the first experiment for 1~s segments is 81.47~\% for the within-trial test set, while the average accuracy of the extra held-out test set does not exceed 58.60~\%. The average accuracy of the second experiment is more consistent, with an average of 62.54~\% for the first test set and an average of 59.67~\% for the extra held-out set, showing the need for a between-trial split when applying deep learning models to the auditory attention paradigm.\\
We also observed (data not shown) that using the between-trial split led to a high variance across trials in the evaluation performance, which indicates that the model often overfits to trial-specific information. In \cite{Vandecappelle2021}, the analysis was performed on a leave-one-story-and-speaker-out cross-validation to avoid overfitting to speakers and stories, which only allowed two possible splits due to the limitations of the dataset. Apparently, these two splits coincidentally led to relatively good results, while other splits could lead to much lower performances as demonstrated in this study.


Overall, out of 13 articles gathered with multiple sound source paradigms, only 4 performed a between-trial split, which as shown in this section, is needed to avoid obtaining a biased model performance.

\begin{table*}[htbp!]
\centering
 \caption{Example division for the Das2019 dataset for 1 subject. Between subjects, the attended direction is alternated.}
 \label{tab:Das2019_dataset}

\hspace{-0cm}\begin{threeparttable}
\vspace{0cm}\begin{tabular}{@{}llll@{}}\toprule

   Trial& Left stimulus & Right stimulus & Attended side  \\ \midrule

      1 & Story1, part1 & Story2, part1 & Left \\
      2 & Story2, part2 & Story1, part2 & Right \\
      3 & Story3, part1 & Story3, part1 & Left \\
      4 & Story4, part2 & Story4, part2 & Right \\
      5 & Story2, part1 & Story1, part1 & Left \\
      6 & Story1, part2 & Story2, part2 & Right \\
      7 & Story4, part1 & Story3, part1 & Left \\
      8 & Story3, part2 & Story4, part2 & Right \\
      9-20 & All stories first 2 min &  All stories first 2 min & Alternate\\\bottomrule
\end{tabular}
\end{threeparttable}
\end{table*}

\subsection{Benchmarking model evaluation using public datasets} \label{public_dataset}

Publicly available datasets include (1) for multiple speech sources (AAD): (\cite{das_neetha_2019_3377911, fuglsang2017}), and (2) for a single speech source:  \cite{fuglsang2017}, \cite{BRODERICK2018803}, \cite{weissbart_hugo_2022_7086168}, and . While we are grateful to the authors for making this data available, unfortunately,  as EEG data collection is expensive and time-consuming, the total amount of data up to now has remained relatively small in the context of deep learning: ~60 hours for AAD, and ~30 hours for single sound source paradigm. 
The lack of a larger public dataset makes it difficult to benchmark the models. Moreover, training and evaluating on a specific dataset can result in overfitting and lack of generalizability.
We recently made available a dataset \citep{K3VSND2023} of 85 subjects, with a total of approximately 170 hours of data. While this remains a very small dataset compared to those available in the fields of automatic image and speech recognition, we hope that it paves the way towards standardized benchmarking such as demonstrated in the recent IEEE Auditory EEG Challenge \citep{Bollens_2023_challenge}.\\

\noindent A potential point of improvement for most of the papers from this review is to \emph{additionally} evaluate the developed architectures on multiple datasets recorded with various EEG devices (e.g., different number and location of electrodes) and experimental set-ups (e.g., different signal-to-noise ratios, inserted phones or speakers).
To illustrate good practices, some generalization experiments were conducted in \cite{Accou2023}: the authors trained a model on their dataset and they evaluated it on a publicly available dataset \citep{fuglsang2017}.\\
Among articles gathered in this study, only 9 out of 29 \citep{Monesi2020, Monesi2021INTERSPEECH, Accou2021ModelingTR, Accou2021PredictingSI, Bollens2021, su2021auditory, kuruvila2021extracting, Vandecappelle2021, Puffay2023.03.30.534911} involved the use of a publicly available dataset, and only one attempted to evaluate generalization to another dataset \citep{thornton2022robust}. Ideally, we recommend to evaluate trained models on multiple publicly available datasets, to ensure their generalization capabilities.

\noindent Considering the above-stated issues, the solution to share data publicly seems straightforward. However, sharing EEG data is complicated due to their biological nature. In many countries, the subjects have to agree explicitly to their data being shared (anonymized/pseudonimyzed) in a publicly available dataset.\\

\noindent Therefore, we encourage the research groups to work towards establishing a common dataset to facilitate model comparison. 
This will be a huge time gain and be a good control for possible pitfalls in recordings, preprocessing, or model evaluation. 
As a comparison, most  deep learning models in ASR are evaluated on shared datasets (e.g., Librispeech ASR corpus from \cite{Librispeech}) and with common error measures such as word error rate. 

\subsection{Interpretation of correlation scores}\label{correlation_score}

When decoding continuous speech features such as the envelope from EEG, decoding quality is often estimated by correlating the reconstructed speech envelope with the presented stimulus envelope. In several papers we collected \citep{Katthi2021DCA, Katthi2021DeepMC, thornton2022robust, Sakthi2019}, a correlation metric is reported as a metric of the performance of the model being used. While correlation metrics are important for interpretation and possible applications (e.g. hearing tests), they depend on the training, evaluation and architecture of a model, the experimental paradigm, and the nature, quality, size and preprocessing of the datasets used. 

Standard statistical tests for correlations are ill-equipped to deal with non-independent sample data, such as (low-pass filtered) EEG and speech envelopes \citep{COMBRISSON2015126, Crosse_review}, as estimated correlations between distant segments can be high by chance. Therefore, an appropriate null distribution has to be constructed to detect whether a model can effectively use neural data to decode speech from EEG (or predict EEG channels from a speech feature). For the encoding/decoding case, \cite{Crosse_review} proposed to use a permutation test using randomly (circular) shifted versions of the predicted data with regards to the actual data to estimate the null distribution. Among all the papers we gathered performing a R/P task, only \cite{thornton2022robust} used this method. The percentiles of this null distribution can be used to measure the significance of the results. In the match-mismatch setup \citep{DeCheveigne2021}, the null distribution is implicitly modeled by the distances between multiple mismatched (transformed) EEG-stimulus pairs when calculating the sensitivity index or match-mismatch accuracy. The typical methods to select mismatched pairs are similar to the methods to construct a null distribution, i.e., permutation test of circular shifts or swapping the ground truth between trials \citep{Crosse_review}.

\begin{figure}
    \centering
    \includegraphics[width=\linewidth]{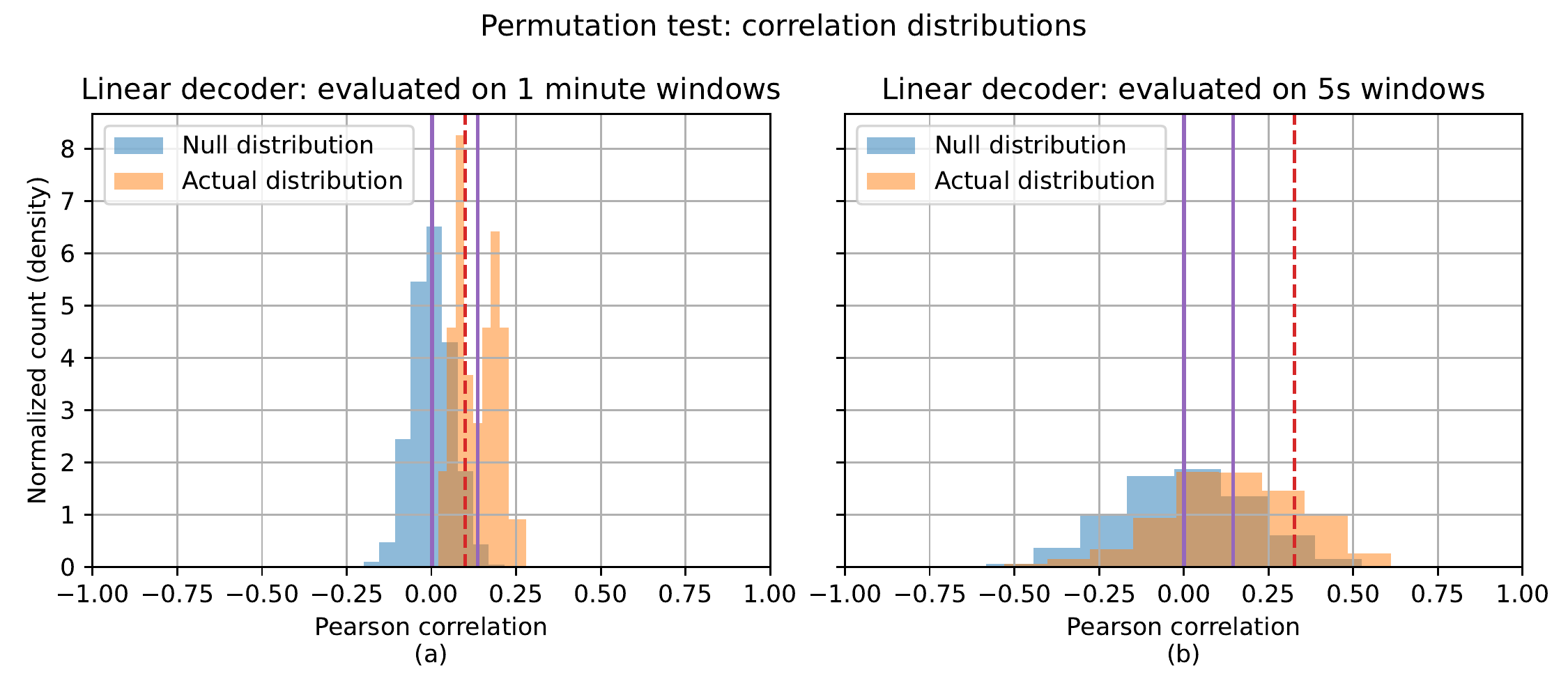}
    \caption{Null and actual (prediction) distributions for a linear decoder with a 250~ms integration window, trained in 6-fold cross validation. For both (a) and (b) EEG and speech data was filtered between 0.5-4Hz for training and evaluation. For (a), evaluation was performed on 1 minute windows with 80\% overlap. For (b), evaluation was performed on 5 second windows with 80\% overlap. The purple lines represent the mean of the null and actual distribution respectively. The red striped line represents the 95-percentile of the null distribution. Note that while the same model is used in (a) and (b), and the mean correlation score in (b) is higher than (a) (0.146 vs. 0.137 respectively), only (a) is statistically significant. }
    \label{fig:corr_distributions}
\end{figure}

To illustrate the pitfalls of using correlation as a performance metric, we trained a linear decoder with a 250~ms integration window on data of a single (representative) recording of the 48-subject dataset in 6-fold cross-validation. Both EEG and speech envelope data were filtered between 0.5-4 Hz with an eighth order Butterworth filter. The decoder was evaluated on 1 minute windows and 5 second windows with 80\% overlap. The null distribution was constructed using 100 permutations of circular shifted speech envelopes, similar to the approach suggested by \cite{Crosse_review}. The results are shown in Figure \ref{fig:corr_distributions} (a). The mean of the predicted correlation scores (0.137 Pearson correlation) for 1 minute windows was greater than the 95th percentile of the null distribution (0.099 Pearson correlation), showing significance at $\alpha$=0.05. While the mean of the predicted correlation scores for 5 second windows was higher than for 1 minute windows (0.146 vs. 0.137 Pearson correlation respectively), comparison with the null distribution shows that the model fails to detect neural tracking in this case (due to the increased variance of the distributions when shorter window sizes are used).

Using the permutation of random circular shifts has a few drawbacks. As mentioned in the study of \cite{Crosse_review}, discontinuities might appear at the end/beginning `of the recording, possibly leading to an inappropriate null distribution. If the recording is sufficiently long however, this risk decreases. Secondly, sufficient permutations have to be computed to obtain an accurate estimate of the percentile of the null distribution. As an alternative, phase scrambling can be used (where the signal is transformed with a frequency transform, the phase is scrambled/randomized and then transformed back into the time domain). Note that while phase scrambling preserves the autocorrelation function, it can result in a slightly more optimistic null distribution for models that can effectively predict phase information (e.g. a decoder that randomly predicts envelope segments).\\

When comparing models based on prediction quality, the choice of preprocessing techniques and datasets should be taken into account. For example, studies commonly filter EEG data into separate bands (e.g., delta band [0.5-4~Hz], theta band [4-8~Hz], etc.) These bands have been linked to different processing stages (e.g., \cite{Etard2019}). When filtering data, caution has to be taken when filtering the target signal (i.e., EEG in forward models, speech features in backward models), as this directly influences the difficulty of the task (e.g., a narrowly bandpassed low-frequency target signal is easier to predict than a broadband target signal), possibly making the task trivial. This also complicates using correlation scores as a metric for general model performance, as some models might perform well using broadband EEG/stimuli features (e.g., \cite{Accou2021PredictingSI}), while others might benefit from more narrowband features (e.g., linear decoders \cite{Vanthornhout2018}). Finally, auditory EEG datasets are often recorded with varying equipment, varying methodologies and different languages of both stimuli and listeners, which can influence the obtained correlation scores and thus make correlation scores unfit for comparison of model performance across datasets.

Our recommendations are as follows: Firstly, construct an appropriate null distribution for each experimental result, and compare it to the correlations between the predicted and original signal. Secondly, when comparing models based on correlation scores of predictions, one must be aware of the influence of external factors (preprocessing, dataset choice, training/evaluation paradigm,...) on the obtained correlation values and interpret the obtained correlation scores with caution. 

We also identified studies that used MSE and MCD as a reconstruction (or prediction) performance metric \citep{Krishna2020, KrishnaEUSIPCO2020, Krishna2021NER}. While MSE is equivalent to Pearson correlation mathematically for normalized vectors, we recommend that researchers provide multiple metrics including Pearson correlation to enable straightforward comparison with other studies. A recent study also implemented a model predicting the EEG signal from speech for both a match and a mismatch segment, in order to get an accuracy value from a forward model \citep{Puffay2023.03.30.534911}, opening the path for a mapping between evaluation metrics.\\

\subsection{Model generalization to unseen subjects}\label{SI_model}

Subject-specific models sometimes have an performance advantage over subject-independent models as they can be fine-tuned to idiosyncrasies of a given subject and are not required to generalize to other subjects. However, subject-independent models are particularly attractive as they do not require training data for new subjects and much larger datasets an can be used to train them. \\
Across subjects, the EEG cap placement can vary and so does the brain activity. Training models on multiple subjects enables the model to learn these differences. That remark also applies to different EEG systems with different densities and locations of electrodes, or experiment protocols.\\

\begin{figure}
    \centering
    \includegraphics[width=\linewidth]{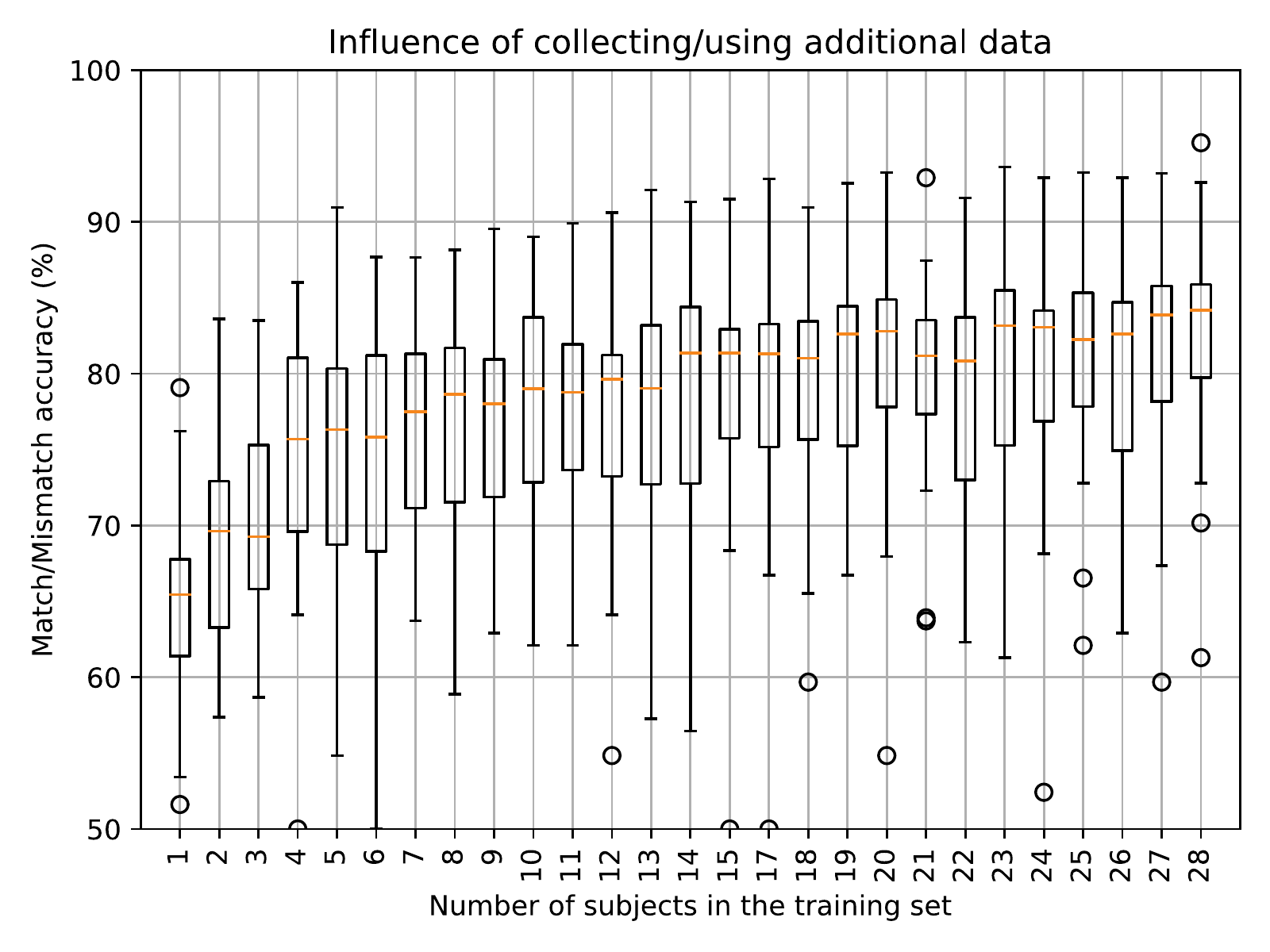}
    \caption{The LSTM model of \cite{Monesi2020} trained on 1-28 subjects of the 48 subject dataset, and evaluated on the test-set of the remaining 20 subjects. Each point in the boxplot corresponds to the match-mismatch accuracy for one subject, averaged over segments. }
    \label{fig:lstm_learning_curve}
\end{figure}

The performance of subject-independent models, especially on subjects not seen during training, depends on the training data. To illustrate this, the LSTM model of \cite{Monesi2020} is trained on 1 up to 28 subjects of the 48-subject dataset, and evaluated on the test set of the 20 remaining subjects. The results are displayed in Figure \ref{fig:lstm_learning_curve}. With this analysis, we show that given the model and the collected dataset, the performance seems to reach a plateau, as the standard deviation of the last 10 medians (from subject number 18 until subject number 28) is less than 1\%.\\

Among the 29 studies we gathered, 7 used a subject-independent training paradigm. If one has a limited amount of data per subject, we recommend to use subject-independent training. One can still fine-tune a subject-independent model (i.e. train a subject-independent model on all subjects, keep its weights and train it on the subject of interest before evaluation) to boost its performance on a given subject.  Please note that fine-tuning will likely be more efficient if the subject fine-tuned on belongs to a similar group to subjects used for prior training (e.g., healthy young normal-hearing).

\subsection{Negative samples selection in MM task}\label{MM_negative_sample}

\subsubsection{Training phase: suggestions to avoid learning spurious cues with the MM task.}

\begin{figure}[htbp!]

\centering
	
	\begin{minipage}[b]{1\linewidth}
		\centering
		\centerline{\includegraphics[width=\textwidth]{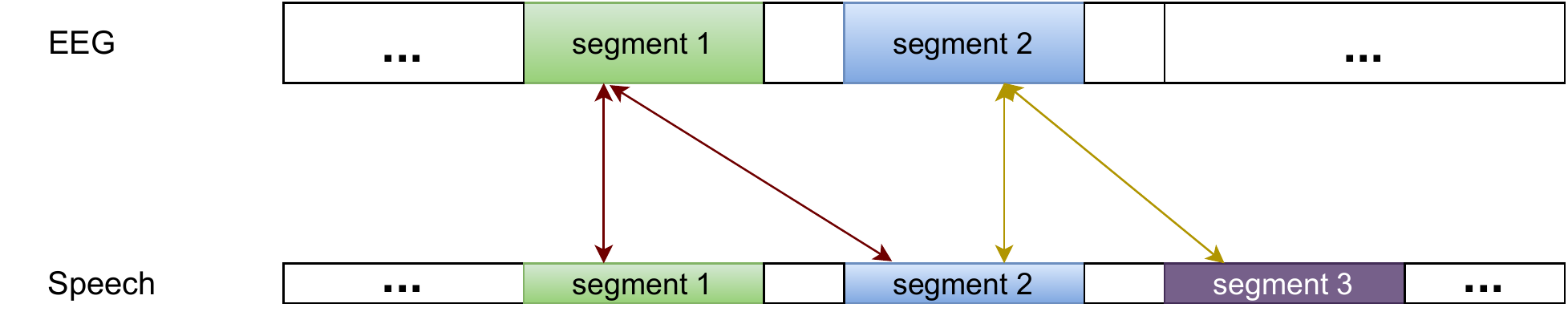}}
	\end{minipage}

	\caption{Illustration of the choice of mismatched segments in the match-mismatch task. Two important points to consider: 1. The mismatched segments are taken from the same sequence. 2. The mismatched segments are also matched segments (and vice versa) depending on the EEG segment. For example, segment 2 speech is the mismatched candidate with segment1 EEG but it will be a matched speech candidate with segment2 EEG.}
	\label{fig:mismatch_samples}
\end{figure}

When training models on the MM task, the choice of the mismatched segments (negative samples) is important to make sure the model can generalize well later in the evaluation phase. The negative samples should be what is called ``hard negatives" in the contrastive learning literature \citep{Oord2018RepresentationLW}. The negative samples should be challenging enough (i.e., with a distribution sufficiently similar to the positive samples) such that it forces the model to learn to relate EEG to the positive speech samples (here the matched segments) rather than only learning to distinguish between positive and negative samples.\\
Training can be performed with one \citep[e.g.,][]{Monesi2020} of multiple negative samples \citep[e.g.,][]{DeCheveigne2021}. As long as the evaluation is robust to biases, all training approaches are worth being tried, however, for the sake of simplicity, we here use a single negative sample for training.\\
To train a model on an MM task that can relate EEG to speech, we give three suggestions to facilitate generalization later in the evaluation phase: (1) select a mismatched segment temporally proximal to the matched segment (``hard negative"); (2) each speech segment should be labeled once as matched and once as mismatched (see Figure \ref{fig:mismatch_samples}), to avoid the model learning a spurious speech segment to label association, without using information from the EEG segment; and (3) sample mismatched segments from the same speech stimulus (i.e., story) as the the matched segments. We encountered issues while not following these suggestions and demonstrate with the three experiments below how they can impact the evaluation performance.\\

\begin{figure}[htbp!]
\centering

\begin{subfigure}{0.9\textwidth}
    \centering
    \includegraphics[width=\textwidth]{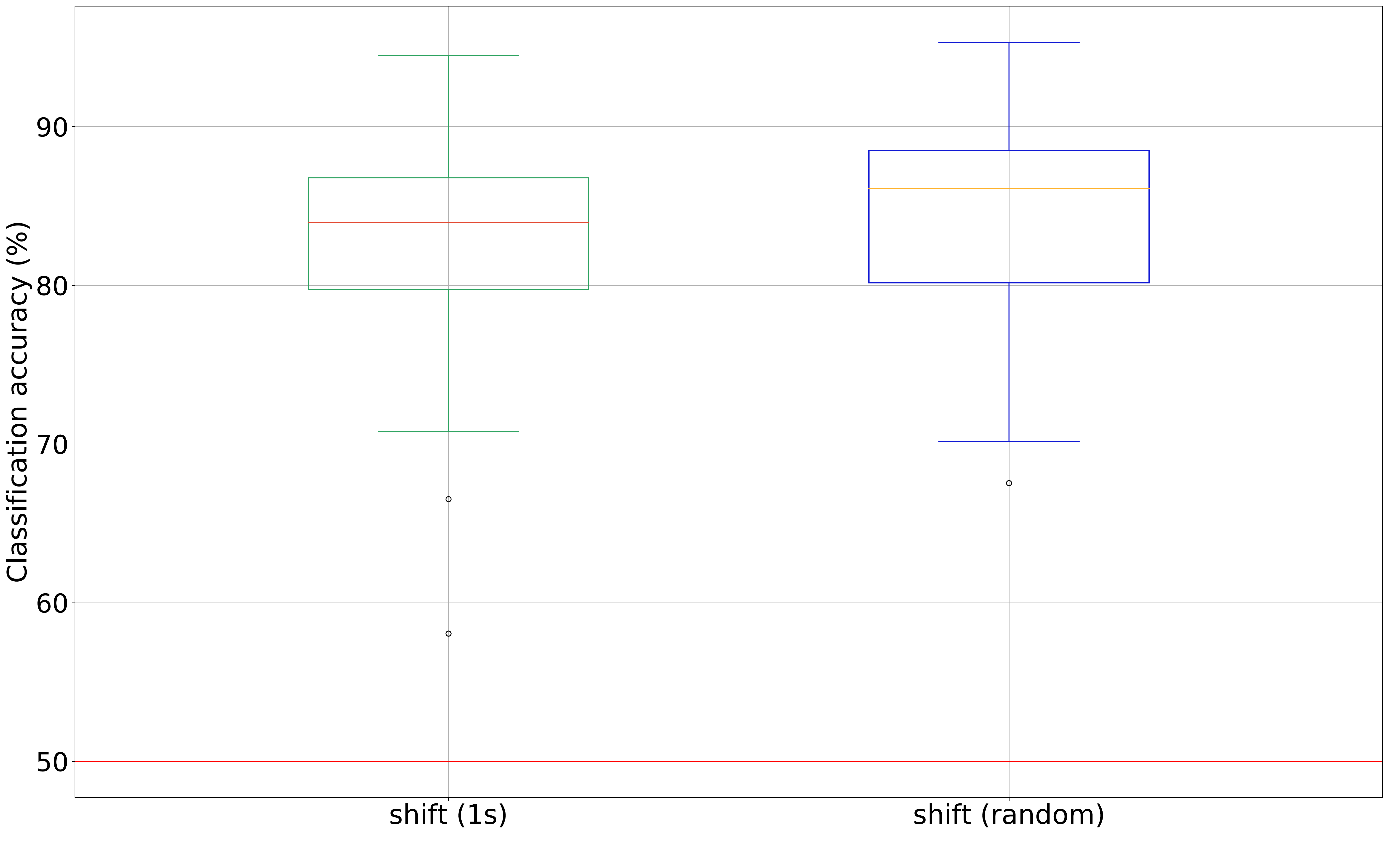}
    \caption{Match-mismatch accuracy of a LSTM model as a function of the mismatched shift.}
    \label{fig:shift_exp}
\end{subfigure}
\hfill
\begin{subfigure}{0.9\textwidth}
    \centering
    \includegraphics[width=\textwidth]{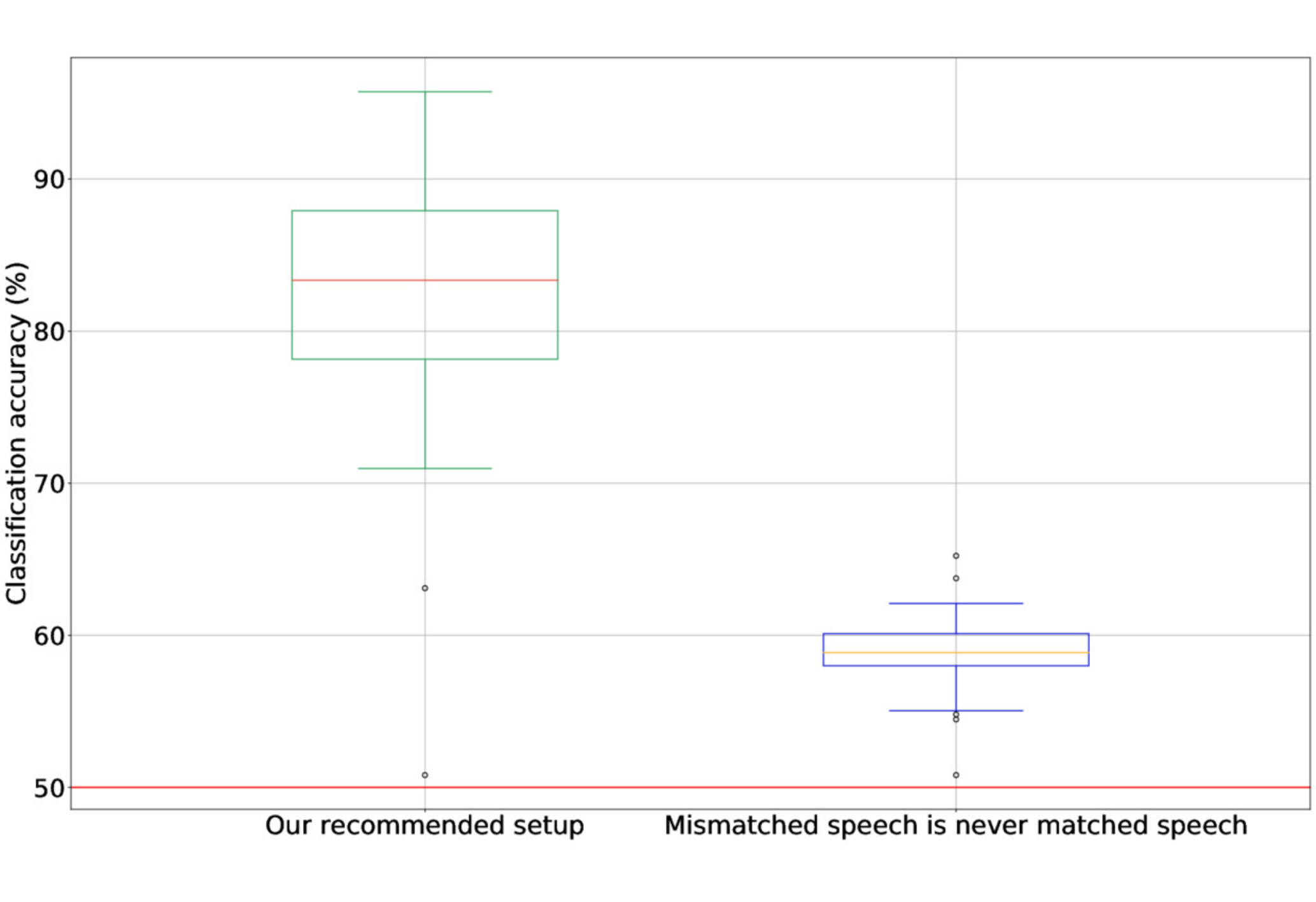}
    \caption{Classification accuracy of the LSTM-based model in the match-mismatch task. Box plots are shown over 48 subjects.}
    \label{fig:unique_mismatch}
\end{subfigure}

\caption{\textbf{Match-mismatch accuracy following different training set-ups.} (a) Match-mismatch accuracy of a LSTM model as a function of the mismatched shift. Our LSTM model was trained on the training and validation set of our 48 subjects dataset with a shift of 1~s between the matched and mismatched segments. This model was evaluated on the test set of the same subjects with a 1~s shift and a randomly picked shift (between 1~s and 20~s) per subject. ; (b) Classification accuracy of the LSTM-based model in the match-mismatch task. Box plots are shown over 48 subjects. Our recommended setup: our proposed training setup where a mismatched segment is also a matched segment with another EEG segment. Mismatched speech is never matched speech: in this setup, a mismatched speech segment will never be exactly a matched segment. As a result, some speech segments will only be matched and others only mismatched.}

\end{figure}

To support the choice of a temporally-proximal mismatched segment (i.e., suggestion 1) as a robust hard negative, we conducted the following experiment: we train our LSTM model on 48 subjects from the dataset with matched and mismatched segments selected with a 1~s shift. Please note that this shift can also be seen as hyperparameter to optimize for a given model and dataset.
We then evaluate our model under two conditions. First on a test set generated with the same 1~s shift for the 48 subjects, second with a given shift between 1~s and 20~s selected randomly for each subject. We show the results in Figure \ref{fig:shift_exp}.\\
We observe no significant difference in accuracy between the random and the 1~s shift. This result indicates that the model can generalize from the 1~s shift to random shifts and did not simply learn the signature of a fixed shift on the data (which could potentially be present due to serial correlation). In addition, we observe a slight increase of the random shift condition, suggesting that larger shifts facilitates the task for the model.\\

In a  second experiment, we designed our match-mismatched segments in a way that violated suggestion (2) (see Figure \ref{fig:mismatch_samples}) such that the mismatched segments were never exactly matched segments. More specifically, we used 65 time samples (instead of 64) as a space between end of the matched and start of the mismatched segment in combination with using a window shift of 64 time samples (one second).
As a result, matched segments overlapped with mismatched segments but they were never exactly the same. Note that other spacing lengths between end of matched and start of the mismatched segment would also result in mismatched segments never exactly being matched segments with other EEG segments. More generally, if the sum of the window length and the spacing is divisible by the window shift, then mismatched segments will also appear as matched segments (our recommended setup for training).\\
We used the 48-subjects dataset where subjects listened to 8 stories. Each recording was split into training, validation, and test sets using 80\%, 10\%, and 10\% ratios, respectively. The training set comprised 40\% from the start and 40\% from end of the recording and the remaining 20\% was further split into validation and test sets. As shown in Figure \ref{fig:unique_mismatch}, the model performs poorly when mismatched segments are never matched segments (i.e, with a 65 samples spacing). Note that the dataset has only around 2.5 hours of unique speech. As a result, the model learned to remember the matched and the mismatched speech segments (of the training set, the training accuracy is around 90\%) instead of relating them to EEG, which is presumably a harder task.\\

In a third experiment, we trained our model with matched and mismatched speech candidates from different stimuli (stories), to simulate what happens when violating suggestion (3). In that case, the stimulus used for the mismatched segment was randomly chosen from a set of 7 stories available for the subject. We compare the performance with that obtained when training the model with the matched and mismatched segments from the same stimulus (i.e., our default scenario).\\
When violating suggestion (3) during training, the model does not generalize well (52\% classification accuracy) to unseen stimuli and also performs poorly (53\% classification accuracy) when evaluated on the default scenario.\\
On the other hand, we observe that the model trained with the default scenario performs well on unseen data (84\% classification accuracy) and also with matched and mismatched segments extracted from different stimuli (84\% classification accuracy). This suggests that the model has learnt to find the relationship between EEG and the stimulus using the default scenario, and not when violating suggestion (3).\\

\subsubsection{Evaluation phase: recommendations for a robust measure.}

Once the model is trained on the MM task, it is crucial to evaluate the model in a robust manner to establish, without any bias, whether it is able to generalize on unseen data. We suggest three methods while stating their advantages and drawbacks.\\
We consider that in the MM task, the model classifies a segment as matched most of the time using a given correspondence measure between the EEG segment and the matched (and mismatched) segment(s). Examples of correspondence measures are the cosine similarity \citep{Monesi2020, Accou2021ModelingTR, Puffay2022} and the Euclidean distance \citep{DeCheveigne2021}. For clarity, we define the correspondence measure between a match candidate and the EEG segment as $d_{m}$, and the correspondence measure between a mismatched candidate and the EEG segment as $d_{mm}$. The three methods are as follows:

\begin{enumerate}
    \item for each matched candidate, compare the mean across all the possible $d_{mm}$ with $d_{m}$ \citep{DeCheveigne2021} \footnote{In case no correspondence measure is used (e.g., direct nonlinear classification), one might use all the mismatched segments separately and average the resulting accuracy, rather than taking the mean distance.}
    \item for each matched candidate, compare $d_{mm}$ of an arbitrary selected mismatched segment and $d_{m}$
    \item for all matched candidates, compare $d_{mm}$ of a mismatched segment selected with a fixed delay and $d_{m}$. Then verify that the results are not better than those from (ii).
\end{enumerate}

Those three methods have their pros and cons: (i) presents the advantage of limiting the variability of the decision criterion, while it is more computationally expensive than (ii) and (iii) due to the calculation of the correspondence measure for each possible mismatch.
(ii) has more variability than (i) and (iii) as the temporal proximity with the mismatched segment varies from one matched segment to the next one.
Finally, (iii) is less time-consuming than (i), and have less variability than (ii). However, it does not ensure the generalization of the model to other shifts, which might hide a bias in the evaluation performance, hence the need for a verification than it does not perform better than (ii).

\section{Conclusions}


We gave an overview of the methods to relate EEG to continuous speech using deep learning models. Although many different network types have been implemented across studies, there is no consensus on which one gives the best performance. Performance is difficult to compare across studies as most research groups use their own dataset (e.g., EEG device, subjects) and training paradigms. 
As we suspected many cases of overfitting, we suggested guidelines to make the performance evaluation less biased, and more comparable across studies. \\

The first point addressed the importance of the training, validation and test set selection. We demonstrated with an experiment that in multiple speech sources paradigms, the split must not be done within trials (i.e., the subjects have to pay attention to one of the two speakers during a defined amount of time) but between them. 
Some studies we reviewed have done such a split and show implausibly high decoding accuracies \citep[e.g.,]{lu2021auditory, su2021auditory}, possibly remembering each trial's label when the split is done within trials.\\

We then addressed the need to use and share public datasets to encourage researchers to improve models and have a common general evaluation benchmark to do so. Gathering diverse data is also necessary to make models more generalizable across devices or experimental setups. We propose to proceed similarly to ASR and computer vision research by gathering large and diverse public datasets rather than working separately on small personal datasets.\\

While correlation metrics are important for interpretation and possible applications (e.g. hearing tests), they depend on the training, evaluation and architecture of a model, the experimental paradigm, and the nature, quality, size and preprocessing of the datasets used. It is necessary to construct an appropriate null distribution for each experimental result to see if a model performs significantly better than chance. When comparing models based on correlation scores of predictions, researchers should be aware of the influence of external factors (preprocessing, dataset choice, training/evaluation paradigm,...) on the obtained correlation values and interpret the obtained correlation scores with caution.

Subject-independent models are very convenient because, when trained on a sufficient amount of data, they can cope with dataset diversity due to, e.g., EEG devices, protocols, brain anatomy or speech content. Although in certain cases (e.g., hearing aid device), an individual's good performance prevails over an ability to generalize, deep learning models require lots of data, which is not clinically ideal to collect from individual subject. We therefore recommend to use subject-independent models when the amount of data is limited.\\
For practical applications, we need deep learning models to generalize and researchers to test their ability to do so, notably by evaluating models on other datasets or ensuring they were trained on enough data to reach their optimal performance. As an example experiment, we characterized an LSTM model's performance as a function of the number of subjects included in the training.\\

Finally, we underline the importance of the negative sample selection in the training phase of a match-mismatch task, and of the implementation of a robust evaluation method. Our suggestions for the training phase will possibly improve the ability of models to generalize to unseen data, while our recommendations for the evaluation are a robust indicator of the ability of a model to measure neural tracking without using biases from the paradigm itself.\\

Hence, two important characteristics of the negative sample selection are that the mismatched segment is taken from the same speech segment and that each mismatch speech segment is also a matched speech segment with another EEG segment. These two points constrain the model to use the EEG data provided to the model, ensuring the model cannot find the matched segment solely from the speech data.\\


\section{Acknowledgements}

Funding was provided by the KU Leuven Special Research Fund C24/18/099 (C2 project to Tom Francart and Hugo Van hamme), FWO fellowships to Bernd Accou (1S89622N), Corentin Puffay (1S49823N), Lies Bollens (1SB1423N) and Jonas Vanthornhout (1290821N).

\bibliographystyle{plainnat}
\newcommand{\newblock}{}
\bibliography{reference.bib}

\appendix

\section{Artificial neural network architecture types}\label{appendix_architecture}

In Section \ref{Review}, studies refer to different network architecture types which are introduced in this section. 

\subsubsection{Fully-connected neural network (FCNN)}

A fully-connected neural network (FCNN) is composed of fully-connected layers in a neural network where all the inputs $x_{j}$ from one layer are connected to every unit $h_{i}$ of the next layer ($n$ in total). The output $h_{i}(\textbf{x})$ of a given unit with index $i$ is defined in Equation \ref{eq:FC} below. With $f$ a nonlinear transformation, $w_{i}$ the weight given to input $x_{j}$ and $b_{i}$ the corresponding bias.

\begin{equation}
    h_{i}(\textbf{x}) = f\left(\sum_{j=1}^{n} w_{ij}x_{j}+b_{i}\right)
    \label{eq:FC}
\end{equation}

\subsubsection{Radial basis function (RBF) network}

\begin{figure}[htpb!]
    \centering
    \includegraphics[width=0.5\textwidth]{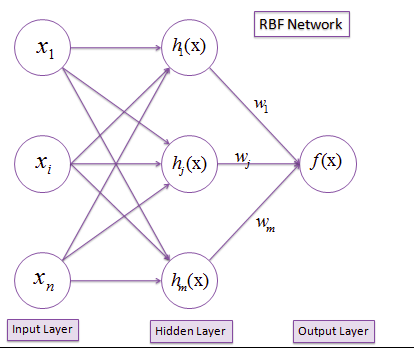}
    \caption{Radial basis function (RBF) network. $x_{i}$ is feature $i$ of a given input, $h_{i}$ is the Gaussian activation function and $w_{i}$ is the weight attributed to $h_{i}$.}
    \label{fig:RBF}
\end{figure}

An RBF network is an artificial neural network composed of 3 layers  (input, hidden, and output) and uses RBF as activation functions. A typical architecture is depicted in Figure \ref{fig:RBF}. All vectors will be depicted in bold and matrices with capital letters in the following descriptions.

\noindent A radial basis function $h_{i}$ is a Gaussian function defined in Equation \ref{eq:RBF} below. It depends on the Euclidean distance of the input $\textbf{x}$ to the center (or average) $c_{i}$ of each input and the radius (or standard deviation) $r_i$.

\begin{equation}
    h_{i}(\textbf{x}) = e^{-\frac{(\textbf{x}-c_{i})^{2}}{2r_{i}^{2}}}
    \label{eq:RBF}
\end{equation}

\noindent The output $f(\textbf{x})$ of the network is a linear combination of RBFs of the inputs and neuron parameters (see Equation \ref{eq:RBF_output} below).

\begin{equation}
    f(\textbf{x}) = \sum_{i=1}^{n} w_{i}h_{i}(\textbf{x})
    \label{eq:RBF_output}
\end{equation}

\subsubsection{General regression neural network (GRNN)}

A GRNN is an RBF network with a slightly different hidden layer. 
The activation function is defined in Equation \ref{eq:GRNN} and the output $f(\textbf{x})$ in Equation \ref{eq:GRNN_output}. 

\begin{equation}
    h_{i}(\textbf{x}) = e^{-\frac{(\textbf{x}-c_{i})^{2}}{2r_{i}^{2}}}
    \label{eq:GRNN}
\end{equation}

\begin{equation}
    f(\textbf{x}) = \frac{\sum_{i=1}^{n} w_{i}h_{i}(\textbf{x})}{\sum_{i=1}^{n} h_{i}(\textbf{x})}
    \label{eq:GRNN_output}
\end{equation}

\subsubsection{Convolutional neural network (CNN)}

A convolutional neural network (CNN) is a deep learning algorithm that takes an input, and apply a sliding filter to it. A CNN contains convolutional layers, which slide a filter over the input.\\
A $N\times N$ input followed by an $m\times k$ filter will output an $(N-m+1)\times (N-k+1)$ matrix.
The output $y_{ij}$ of a convolutional layer taking $\textbf{x}$ as input is defined in Equation \ref{eq:CNN_output} below. $(i, j)$ being the position of a given element in the $n \times n$ input, $a$ and $b$ being the shift applied to dimensions 1 and 2 of the input respectively, $w_{ab}$ the weight corresponding to the element shifted by $a$ and $b$ and $f$ a given nonlinear transformation.

\begin{equation}
    y_{ij} = f\left(\sum_{a=0}^{m-1}\sum_{b=0}^{k-1} x_{ab}h_{(i-a)(j-b)}\right)
    \label{eq:CNN_output}
\end{equation}

\noindent A CNN often ends with at least a fully-connected layer to compile the data extracted previously by convolutional layers to form the final output.

\subsubsection{Long-short term memory (LSTM) based neural network}

Unlike standard feed-forward neural networks, long-short-term memory (LSTM) has feedback connections. A common LSTM unit is composed of a cell, an input gate, a forget gate, and an output gate. The cell remembers values over arbitrary time intervals and the three gates regulate the flow of information into and out of the cell.\\

\noindent The equations for the gates in LSTM are: 

\begin{equation}
    i_{t} = \sigma(w_{i}[\textbf{h}_{t-1} , \textbf{x}_{t}]+b_{i})
    \label{eq:input_gate}
\end{equation}

\begin{equation}
    f_{t} = \sigma(w_{f}[\textbf{h}_{t-1}, \textbf{x}_{t}]+b_{f})
    \label{eq:forget_gate}
\end{equation}

\begin{equation}
    o_{t} = \sigma(w_{o}[\textbf{h}_{t-1}, \textbf{x}_{t}]+b_{o})
    \label{eq:output_gate}
\end{equation}

where $i_{t}$, $f_{t}$ and $o_{t}$ are input, forget and output gate's equations for time index $t$ respectively. These gates all apply a sigmoid to the weighted sum of a combination of the previous hidden state $\textbf{h}_{t-1}$ and the current input $\textbf{x}_{t}$. Each gate has its own weight and bias $w_{x}$ and $b_{x}$, $x=\{i; f; o\}$.\\

\noindent Finally, the candidate cell state $\tilde{\textbf{c}_{t}}$; the cell state $\textbf{c}_{t}$ and the output $\textbf{h}_{t}$ are computed following Equation \ref{eq:cc_state}, \ref{eq:c_state} and \ref{eq:output}.

\begin{equation}
    \tilde{\textbf{c}_{t}} = tanh(w_{c}[\textbf{h}_{t-1}, \textbf{x}_{t}]+b_{c})
    \label{eq:cc_state}
\end{equation}

\begin{equation}
    \textbf{c}_{t} = f_{t}*\textbf{c}_{t-1}+i_{t}*\tilde{\textbf{c}_{t}} 
    \label{eq:c_state}
\end{equation}

\begin{equation}
    \textbf{h}_{t} = o_{t}*tanh(\textbf{c}_{t})
    \label{eq:output}
\end{equation}

\noindent To get the memory vector for the current timestamp ($\textbf{c}_{t}$) the candidate $\tilde{\textbf{c}_{t}}$ is calculated. From the above equation we can see that at any timestamp, our cell state knows that what it needs to forget from the previous state(i.e., $f_{t} * \textbf{c}_{t-1}$) and what it needs to consider from the current timestamp (i.e., $i_{t} * \tilde{\textbf{c}_{t}}$).

\subsubsection{Gated recurrent unit (GRU) based neural network}

Gated recurrent units (GRUs) are a gating mechanism in recurrent neural networks. The GRU is like an LSTM with a forget gate, but has fewer parameters than LSTM, as it lacks an output gate.

\subsubsection{Autoencoders}

\begin{figure}[htpb!]
    \centering
    \includegraphics[width=0.6\textwidth]{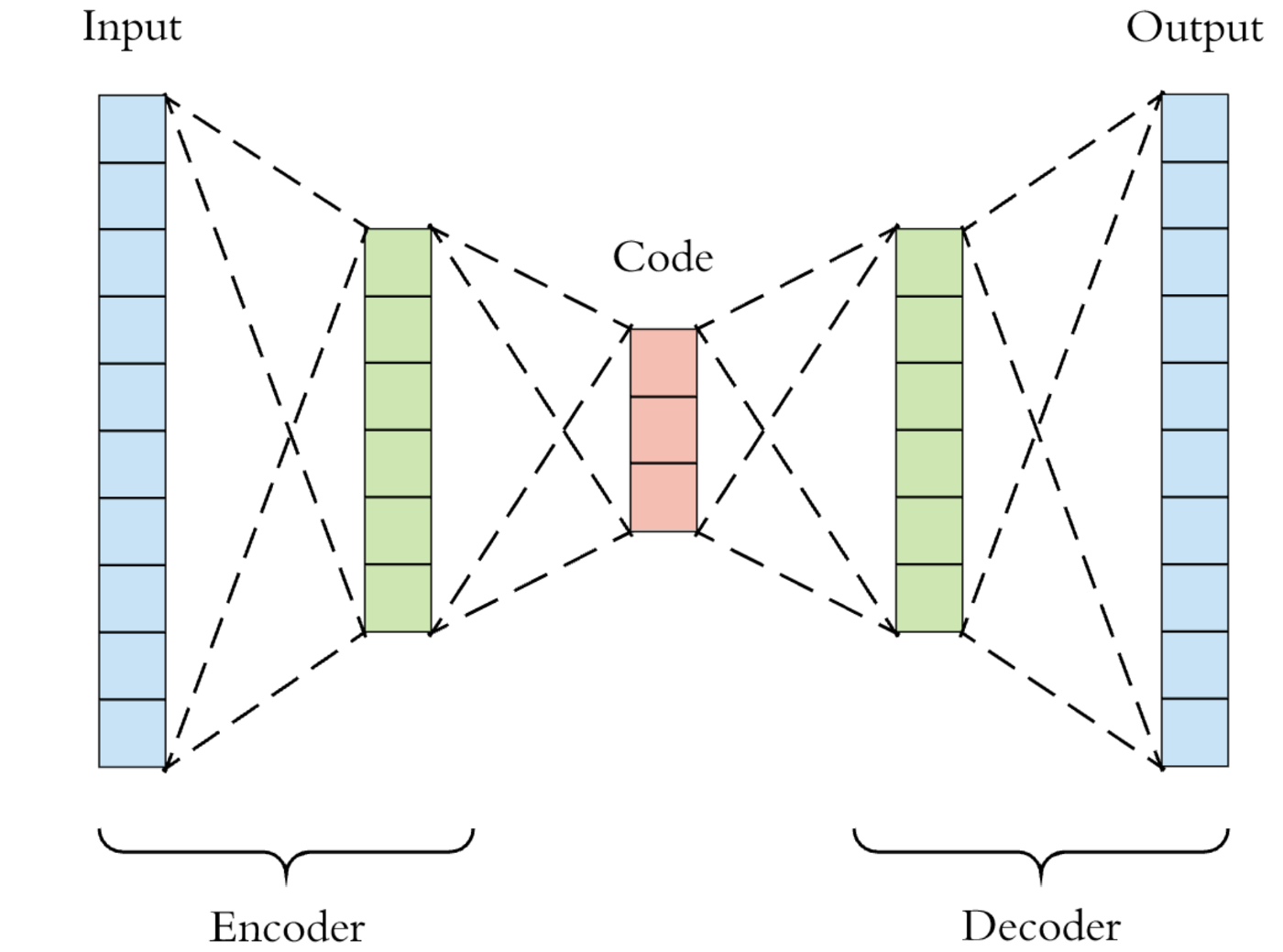}
    \caption{Autoencoder network structure.}
    \label{fig:AE}
\end{figure}

An autoencoder is a type of artificial neural network used to learn efficient encoding of unlabeled data (unsupervised learning). The encoding is validated and refined by attempting to regenerate the input from the encoding. The structure is depicted in Figure \ref{fig:AE}.

Essentially, we split the network into two segments, the encoder, and the decoder.
The encoding network can be represented by the standard neural network function passed through an activation function (see Equation \ref{eq:encoder}), where $\textbf{z}$ is the latent representation, $\sigma$ a non-linear transformation and $W$ a matrix of weigths.

\begin{equation}
    \textbf{z} = \sigma(W\textbf{x}+b)
    \label{eq:encoder}
\end{equation}

\noindent The decoding network can be represented in the same manner, but with different weight, bias, and potentially activation functions (see Equation \ref{eq:decoder}).
The latent dimension $\textbf{z}$ is used to retrieve the input prediction $\textbf{x}'$ and typically the AE is trained on a loss to minimize the distance between $\textbf{x}$ and $\textbf{x}'$.

\begin{equation}
    \textbf{x}' = \sigma'(W'\textbf{z}+b')
    \label{eq:decoder}
\end{equation}

\subsubsection{Transformers}


A transformer neural network is a novel architecture that aims to solve sequence-to-sequence tasks while handling long-range dependencies with ease (which is typically limited in RNNs). It is an architecture based on attention layers introduced in \cite{NIPS2017_3f5ee243}.
Compared to a simple sequence-to-sequence model (e.g., LSTM-based model), the encoder part passes all the hidden states to the decoder rather than solely the previous hidden state. Moreover, the input sequences can be passed in parallel which drastically speeds up the training process.

\section{Summaries of multiple speech source ($N>1$) papers}\label{multiple_source_appendix}

\noindent \textit{A novel technique for identifying attentional selection in a dichotic environment (Shree et al., 2016):}
\cite{shree2016novel} implement a GRNN model to classify left/right attention.
The dataset contains 20 subjects. 10 out of these subjects were instructed to listen to the story played in the left ear, whereas the other 10 had to pay attention to the story played in the right ear. Each subject listened to three trials of 1 minute. The authors used randomized sub-sampling, a method that randomly selects 50\% of the data for training and the other 50\% for testing. It is unclear from the paper whether this split is done per subject or for all subjects. We thus cannot state if the training is subject-specific or subject-independent.\\
A general regression neural network (GRNN) based classifier is proposed, a single-pass architecture that estimates the conditional mean $E[\frac{Y}{X}]$ in order to estimate the most probable values of output $Y$ given input $X$. The model has 4 fully-connected layers. The last layer gives the conditional mean as output.\\ 
The average classification accuracy on a 60~s window of the model is 99.05\%.\\
The classification accuracy is surprisingly high compared to previous work with linear models \citep{Osullivan2015} that obtained 89\%. Linear unsupervised approaches also performed below 90\% on public datasets \citep{geirnaert2021unsupervised}. 
One possible explanation could be that the model associates a subject with a side as each subject either focused on the left or right story.
Another explanation is the small amount of data (3~min) per subject, which considering a network with 4 fully-connected layers might be too little. It is not clear from the paper how many neurons are attributed to each layer, hence we cannot estimate the number of parameters.
In addition, a major critic to this paper is the lack of information: the training paradigm (subject-specific or subject-independent) is unclear and we do not have information about the model's parameters. However, the two explanations stated above apply in any case.\\

\noindent\textit{Comparison of Two-Talker Attention Decoding from EEG with Nonlinear Neural Networks and Linear Methods (Ciccarelli et al. 2019):}
Often AAD tasks are solved by two subsequent models: an EEG-based speech reconstruction model followed by a model to determine the similarity between the candidate speech streams and the reconstruction, and classify which one was attended. \cite{Ciccarelli2019} introduce an architecture that integrates these two models in one. They evaluate it on a two-competing talker scenario, and compare it to a linear stimulus-reconstruction decoder \citep{Osullivan2014} as well as the neural-network stimulus-reconstruction decoder from \cite{deTaillez2020MachineSpeech} we reviewed.\\
The dataset consists of 11 subjects listening to 40 minutes of two-competing speaker stimulus (four 5-minutes stimuli used once as a distractor, once as an attended speech). 
The model is a CNN-based architecture for direct talker classification. It consists of two convolutional layers, with respectively a kernel of three and one samples. They are followed by a set of four fully connected layers of decreasing size. The training was subject-specific using cross-validation.\\
The CNN classifier approach dramatically outperformed both the neural network and the linear decoder traditional segregated architecture decoding accuracy on a 10~s decision window (81\% vs 62\% vs 66\% respectively).\\
One issue to address is the subject-specific training. The models are not trained with the perspective to generalize across subjects, which is one of the expected improvements from deep learning models over linear models (see Section \ref{Pitfalls}).\\

\noindent\textit{Machine learning for decoding listeners' attention from electroencephalography evoked by continuous speech (de Taillez et al. 2020): }
This is one of the first studies to use neural networks in auditory attention decoding. The authors start from the backward linear model, trying to predict the envelope from EEG, and add non-linearities in a two-competing talker scenario.\\
The dataset contains 20 subjects with a total of 50 minutes of data each. 10 subjects attended solely one speaker, 10 subjects the other one. The whole dataset is divided into 10\%   validation, 10\% testing, and 80\% training data. \\
The model consists of 2 fully-connected (FC) layers . The authors compare different combinations of parameters such as the EEG bandpass filter cutoff frequencies, the length of the training window (or kernel size), the type of cost function (mean squared error or Pearson correlation), the decision window's length and the activation function.\\
The authors report the rate for identifying the attended speaker in terms of bits/minute. We converted that metric into attention decoding accuracy to ease  comparison to other studies. The decoding accuracy obtained with the NN for 60~s, 10~s and 5~s are 97.6\%, 86\% and 79\% respectively. The linear model presented in \citep{Osullivan2015} is outperformed by the neural network.\\
The same stimuli from the two speakers are presented to all subjects. Since the split of the dataset is across subjects, it is likely that the model sees the whole attended (or unattended) stimulus during training. This could lead to overfitting on the stimulus content, hence boosting the model's decoding accuracy. \\


\noindent\textit{Auditory attention tracking states in a cocktail party environment can be decoded by deep convolutional neural networks (Tian and Ma, 2020):}
The authors propose a CNN model with source-spatial feature images (SSFIs) as the input to decode auditory attention tracking states in a cocktail party environment.
The dataset consists of 42 subjects listening to two different Mandarin speakers. Each subject completes 120 trials of 2~s.\\ 
The neural network consists of three convolutional layers. After each convolutional layer, a max-pooling operation is performed. The output of the last max-pooling layer is flattened and fed into a 128-dimensional FC layer, which is fed to the two-dimensional output layer (i.e., representing the probability of the two classes correct or wrong).\\
For the behavioral performance score, they utilize the signed residual time \citep{Maris2012}.
The signed residual time measures the trade-off between response accuracy and task execution speed. Among 42 subjects, 30 with a high signed residual time score are classified as the high behavior performance group (H-group), whereas 12 with low scores are classified as the low behavior performance group (L-group).\\
For both H- and L-groups, the model is trained on all subjects minus one from the same group and evaluated on the remaining one (i.e., inter-subject leave-one-out cross-validation) to obtain the classification accuracy.
Considering the small size of the L-group, a fine-tuning condition is also performed. For each subject from the L-group, the model is pre-trained on the H-group and fine-tuned with an inter-subject leave-one-out cross-validation on the L-group. These models are therefore subject-independent.
The average auditory tracking state classification accuracy (correct or wrong) of the H-group is 80.4\%, while the classification accuracy of the L-group is 69.9\%. 
When fine-tuning the subjects of the L-group on the model of the H-group, in the same inter-subject leave-one-out cross-validation setting, a classification accuracy of 75.2\% is achieved.\\
One caveat here is the small stimulus content (eighteen possible 3-word combinations with common words). The model could possibly identify the EEG response to these words and fail on unseen words (i.e., overfitting).\\

\noindent\textit{Auditory attention detection with EEG channel attention (Su et al., 2021):}
\cite{su2021auditory} propose an AAD system that integrates a neural channel attention mechanism and a convolutional neural network (CNN) classifier.
A publicly available dataset was used \citep{das_neetha_2019_3377911} (we will refer to it as Das2019). The 16 normal-hearing subjects are instructed to attend to one of two competing speakers. Four Dutch short stories (12~min each), narrated by different male speakers, were used as speech stimuli.
The dataset is split into a 60/20/20 train/test/validation scheme for each subject's dataset. 
The authors propose a channel attention mechanism that predicts a channel mask, corresponding to spatial mapping of the EEG electrodes. This channel mask varies per story and subject. To measure the relationship between EEG channels and speech stimuli, the cosine similarity is used. After calculating the cosine similarity, the channel attention model contains two fully connected layers and a softmax layer, to calculate the final channel mask. This channel mask is then applied to the EEG channels.
The model combines the channel attention mechanism with the CNN model proposed by \cite{Vandecappelle2021}.\\
The authors report results for window lengths of $0.1, 0.5, 1$, and $2$ seconds. For 0.1 seconds, the obtained accuracy is $77.2 \%$, for 2 seconds this accuracy goes up to $88.3 \%$.\\
The accuracy obtained on 2~s is very high compared to other studies  (e.g., 72\% for 2~s \citep{kuruvila2021extracting}, 79\% for 5~s \citep{deTaillez2020MachineSpeech}). It may be prudent to evaluate the generalization capabilities of this complex model that has achieved good performance in a subject-specific training by evaluating it within a subject-independent training paradigm. This topic is addressed in Section \ref{SI_model}\\

\noindent\textit{Supervised binaural source separation using auditory attention decoding in realistic scenarios (Zakeri et al., 2021):}
The authors propose a complete pipeline from speech mixture to a denoised signal based on AAD. Their model attempts to separate the attended speaker from the unattended speaker in realistic scenarios, using different signal-to-noise ratios, levels of reverberation times for a simulated room, and different speaker positions.\\
The dataset contains data from 12 subjects, each listening to 48 trials with a length of 54 seconds. These stimuli are split up into 34 seconds train, 6 seconds validation, and 14 seconds of test data. 
The subjects listened to two audio stories.
Stimuli are generated for different SNRs of 4 dB, 9 dB, for clean speech, and for different reverberation times of 0, 0.5, and 1 second. The different spatial configurations of attended and unattended speakers are (-90\degree, 90\degree), (30\degree, 90\degree), (-30\degree, -90\degree) and (-5\degree, 5\degree). 
Both spectral and spatial features are jointly used to train a set of deep neural networks (DNN) responsible for dereverberation and denoising. 
The AAD processing stage takes the EEG as input and then calculates the phase-locking value (PLV), which is then fed to a Bi-LSTM, and detects the attended speaker. Based on the classification scores of the AAD module, different portions of the calculated masks are selected to resynthesize the attended speaker. The Bi-LSTM contains 10 layers with each 100 hidden units.\\
The authors report an average accuracy of the AAD phase of 66.37\% for 1~s up to 84.57\% for 40~s of data.\\
The accuracy obtained is quite low. As a comparison, \cite{su2021auditory} obtained  $77.2 \%$ on 0.1~s, and $88.3\%$ for 2~s. The model was trained on an AAD task, however it was in the meantime utilized to denoise stimuli with low SNRs. This could account for the reduced accuracy compared to other studies that utilized clean speech.\\

\noindent\textit{EEG-based detection of the locus of auditory attention with convolutional neural networks \citep{Vandecappelle2021}:}
A CNN-based model is introduced to classify left or right attention and uses segments of 1-2~s. The Das2019 dataset is used.\\
The proposed model receives an EEG matrix as input, and passes it to a convolutional layer with a ReLU activation function.
An averaging pooling step is then used over the time dimension, reducing each of the time series dimensions to a single number. After the pooling step, there are two fully connected (FC) layers. The first FC layer contains one neuron per time series dimension and is followed by a sigmoid activation function. Finally, the second FC layer contains two (output) neurons.
As a baseline, a linear decoder model was implemented to be evaluated on the same task.\\
The CNN-based model outperformed the linear decoder on a 1~s decision window (81\% vs 58\% respectively). As a comparison, the common spatial pattern approach defined by \cite{GeirnaertCSP} reached 80\%.\\
Although locus of attention decoding is a much simpler task than decoding the stimulus of the attended speaker, the obtained performance on 1~s is higher than in the paper from \cite{zakeri2021supervised} (66\%).\\

\textit{Extracting the auditory attention in a dual-speaker scenario from EEG using a joint CNN-LSTM model (Kuruvila et al., 2021):}
\cite{kuruvila2021extracting} introduce a joint CNN-LSTM model that takes the EEG signals and the spectrogram of the multiple speakers as inputs and classifies the attention to one of the speakers.\\
They use three datasets, two of which are publicly available (i.e., \cite{das_neetha_2019_3377911} and \cite{fuglsang2017}). The third dataset contains data from 27 subjects, listening to German news extracts, amounting to a total of 30 minutes per subject. 
The model uses the spectrogram of the speech stimuli as input to the speech network. Both EEG and spectrogram are first put through a CNN network, then concatenated and put into the last part of the network together. The EEG subnetwork contains 4 convolutional layers. 
The audio subnetwork contains 5 convolutional layers. The EEG output and the two audio outputs are then put into the final network, consisting of a bidirectional LSTM (bi-LSTM) block, followed by a few fully connected layers. 
The dataset is split into a 75/12.5/12.5 train/test/validation scheme between trials.\\
The best performances on unseen data were obtained when training the model on all three datasets together, suggesting that the model might generalize better when training on more data or that heterogeneous data leads to better generalization. For a window of 2, 3, 4, and 5 seconds, the authors report a mean accuracy of respectively $70.9 \%, 73.9\%, 75.2\% $, and $75.5\%$. Although the model contains 400 000 parameters, a pruning analysis shows that sparsity of 40~\% can be achieved without obtaining significantly lower accuracy.\\

\noindent\textit{Auditory attention decoding from electroencephalography based on long short-term memory networks (Lu et al., 2021):}
\cite{lu2021auditory} propose an LSTM-based architecture to decode auditory attention in a competing two-talker scenario. The aim is to investigate whether an end-to-end nonlinear framework can outperform the state-of-the-art linear models on short-decision windows (1~s).\\
21 subjects participated and each listened to two concurrent speech segments of China Nation Radio. Each speech stimulus has a duration of 52~s and the subjects listened to the stimuli twice, alternating their attention in between the presentation. Each stimulus was split into 30 seconds of training data and 22 seconds of testing data. The number of training samples ranges between 28 and 478, depending on the window length.\\
The authors propose a model consisting of 6 LSTM blocks in total, followed by a fully connected layer and a 2-node output layer. First, for both the attended and the unattended streams, the speech envelope and EEG first go through two separate LSTM blocks (four in total). Then, for both attended and unattended streams, the two LSTM's outputs are subtracted element-wise (e.g, for the attended stream $S_{attended} = LSTM_1(Speech) - LSTM_2(EEG)$) and the resulting signal is then put into another LSTM block (two in total). The outputs of both attended and unattended streams are then concatenated and put into a fully-connected layer with 40 hidden nodes, followed by the final output layer which predicts the attended speaker.\\
For a window length of 1 second, the authors report an average accuracy of $96.12\%$, which is an average over 86 1-s testing sample sequences, with a model trained on 118 1-s sample sequences.\\
These accuracies obtained are extremely high for 1-s windows. Considering the complexity of the model (6 LSTM layers with 20 hidden units each) and the very low amount of data per subject (30~s for training, 22~s for testing), it is plausible that the model overfits the data. This study would benefit from testing its generalizability over subjects and different speech content (see Section \ref{Pitfalls}).\\

\noindent\textit{Speaker-independent brain enhanced speech denoising (Hosseini et al., 2021):} 
The Brain Enhanced Speech Denoiser (BESD) is a speech denoiser; it is provided with the EEG and the multi-talker speech signals and reconstructs the attended speaker speech signal.  
The paper is divided into two tasks: one speaker-specific task, during which the attended speaker identity is provided to the model (i.e., the model is trained speaker-specifically). The performance of the BESD model is compared to a classical denoising autoencoder using solely the mixture speech without EEG.
The second task is a speaker-independent denoising task. The model is trained in a different configuration (i.e., different attended speakers) and therefore needs the EEG information to denoise the mixture speech.\\
The dataset consists of 34 subjects, each listening to 30 trials of 1~minute each. Two subject groups were formed; 17 were asked to pay attention to the left speaker, and 17 to the right. For each subject, five trials were assigned to the test set, 2 to the validation set, and 23 to the training set.
The proposed BESD architecture has an autoencoder structure: one encoder for the speech mixture and one for the EEG activity. Each encoder includes three convolutional blocks with a feature-wise linear modulation (FiLM) \citep{Perez_Strub_de_Vries_Dumoulin_Courville_2018}, which an affine transformation applied to the output of the convolutional blocks. A last convolution is applied after these convolutional blocks. The output of both encoders is concatenated in the latent space in a so-called fusion layer. The latter is fed into a decoder block composed of two convolutional blocks similar to the encoder's (filter size 52 and 100 for convolution 1 and 2, respectively). The last layer is a 1D convolution of filter size one followed by a hyperbolic
tangent. 
As the loss function, the authors used a scale-invariant signal to-distortion ratio (SI-SDR) \citep{LeRoux2019SDRH}, that has been shown to perform well as a general-purpose loss function for time-domain speech enhancement \citep{Kolbaek2020}.\\
The BESD model outperformed the classical denoising autoencoder on the speaker-specific denoising task and significantly enhanced the speech compared to the noisy mixtures without having any prior information on the attended speaker. It is also an end-to-end approach in which all the algorithm modules are trained simultaneously. In contrast, two separately trained networks are sometimes used \citep{CEOLINI2020117282} (e.g., one for speaker separation and a second to select the attended speaker and enhance it).\\
Although this technique might be useful for AAD rather than the original objective of EEG-enhanced speech denoising, evaluation on public datasets using common metrics for the AAD might extend the impact of this work (see Section \ref{public_dataset} and \ref{SI_model}).

\textit{Auditory attention decoding from EEG-based mandarin speech envelope reconstruction (Xu et al., 2022):}
The authors introduce an LSTM-based architecture to decode auditory attention using a Mandarin as the stimulus' language. 
21 subjects participated in the study. Each subject listened to the same 40~minutes of speech stimuli, spoken by two different Mandarin speakers. For each subject, the resulting 40~minutes of EEG are split up as follows: randomly per subject, 6 minutes are extracted for training, 2 minutes for validation, and the remaining 32 minutes for testing. This amounts to a total of 126 minutes for training and 672 minutes for testing, with the 40 minutes of stimuli present in both sets.\\
In \cite{xu2022auditory}, the authors propose to use an LSTM-based model that receives EEG as input and has to output the attended envelope. Pearson correlation is used as a loss metric. As far as we can infer, the model uses 10 consecutive LSTM blocks, each containing 64 hidden nodes. There is no mention of the output layer.\\
The authors report that the best working model uses broadband information, 17 channels, and has an average accuracy of 74.29\% for a window length of 0.15s.\\
We here address the issue of randomly selecting the training, validation, and test sets for each subject from the 40-minute recordings in the protocol. There is a non-negligible possibility that over 21 subjects, the whole speech stimulus corpus is seen by the model during training. 
Under this setting, a verification of the generalization to unseen stimuli is missing. We explain potential solutions to this pitfall further in Section \ref{SI_model}.\\

\noindent\textit{Decoding selective auditory attention with EEG using a transformer model (Xu et al., 2022):}
\cite{xu2022decoding} propose a transformer architecture to decode auditory attention in a competing two-talker scenario. The), aim is to investigate whether an end-to-end nonlinear framework can outperform state-of-the-art linear models.\\
The dataset used is the same as \cite{xu2022auditory}.\\
The EEG signal is provided to the model as input to an AAD-based transformer containing an encoder and decoder block. The encoder contains positional encoding, channel attention, and temporal self-attention models. The following decoder takes as input the encoded EEG and reconstructs the speech envelope. The Pearson correlation is computed between the reconstructed left and right speech envelopes. The attended envelope is selected as the one having the highest correlation with the reconstructed envelope.\\
For a 0.15~s window, the authors obtained an average accuracy of 76.35 \%.\\
The issues previously mentioned for \cite{xu2022auditory} apply to this study as well: the model may be overfitted to the stimulus.

\noindent \textit{Robust decoding of the speech envelope from EEG recordings through deep neural networks (Thornton et al. 2022):}
\cite{thornton2022robust} propose a neural network structure to reconstruct the envelope from EEG recordings in different listening conditions. The authors validate their models both on single-speaker datasets and on the competing talker scenario. In this section, we will focus on the competing talker scenario. 
The competing talker dataset contains 18 subjects who listen to a male and a female speaker, who are simultaneously narrating an audiobook. In the first scenario, subjects are instructed to listen to the male speaker, in the second scenario they are instructed to listen to the female speaker. Each listening condition has a total of 10 minutes of data per subject, recorded in 4 different trials. 
As a baseline, the authors use a linear backward model. Two neural networks are compared. The first is an FCNN. A segment of 50-time samples (400~ms) is first flattened and then put through fully connected layers, with each layer containing fewer neurons than the preceding layer. The output of the final layer is a scalar, representing an estimate of the envelope at the onset of the segment. The second network is a CNN inspired by EEGnet \citep{Lawhern_2018} and performs a temporal convolution on the input, followed by a spatial convolution and a depthwise separable convolution. The output is then flattened and reduced to a single scalar output by taking a linear combination. The models are subject-specific. \\
The correlation between the reconstructed envelope and the attended and unattended speaker is calculated and the classifier picks the envelope with the highest correlation as the attended speaker envelope. The classification accuracy is calculated for windows of 2.5 and 10 seconds. Both the CNN and the FCNN offer a clear improvement over the linear model. For a window length of 10 seconds, the CNN achieves around $80\%$ classification accuracy, while the linear model achieves around $68\%$.\\ 
One issue to address in this study is the use of DNNs as subject-specific models. Although better results are obtained with subject-specific models using their dataset in the single-speaker paradigm; with the perspective of gathering more data (e.g., by using public datasets, see Section \ref{public_dataset}), showing results of subject-independent models would have been relevant. The motivations to use subject-independent models is developed in Section \ref{SI_model}.\\

\section{Summaries of single speech sound source (N=1) studies}\label{single_source_appendix}

\textit{Modeling the relationship between acoustic stimulus and EEG with a dilated convolutional neural network (Accou et al., 2020):}\\
\cite{Accou2021ModelingTR} here present a dilated-convolutional network, and compare its performance to a CNN baseline, and to a linear decoder.\\ 
The dataset consists of EEG data from 48 normal-hearing subjects listening to 10 stories (each about 14~min) in Flemish (Dutch). They use the speech envelope as the stimulus feature. The training, validation, and test sets were divided per subject with an 80:10:10 ratio.\\
\noindent This network takes one segment of EEG and two segments (matched and mismatched) of speech and processes them in separate streams. The model uses multiple dilated convolutions as the encoding layers for both speech and EEG streams. The resulting embedded EEG representation is compared with both embedded stimulus representations using cosine similarity.
The decision layer is a sigmoid layer that classifies match-mismatch based on the cosine similarity scores.\\
The model is evaluated using the classification accuracy on the match-mismatch task averaged over subjects. It outperformed a CNN baseline as well as a linear decoder baseline on the same MM task for 5~s and 10~s decision windows (respectively 80\% and 85\%).\\
The same model was used by \cite{Accou2021PredictingSI} to predict speech intelligibility. Being able to derive an objective measure for speech understanding is crucial as behavioral tests are not possible for a part of 
the population.\\ 
We address one critic in \cite{Accou2021ModelingTR, Accou2021PredictingSI}, regarding the classification accuracy estimation quality. In this study, the authors use one mismatched segment (1~s after the end of the matched segment). To get a better estimate of the classification accuracy, an alternative is to take multiple shifted mismatched segments and average the accuracy obtained \citep{DeCheveigne2021}. Considerations about the negative sample(s) selection within the MM task are discussed in Section \ref{Pitfalls}.\\


\noindent\textit{A LSTM-based architecture to relate speech stimulus to EEG (Monesi et al., 2020):}
Compared to \cite{Accou2021ModelingTR, Accou2021PredictingSI}, \cite{Monesi2020} introduced a model integrating LSTM layers (see description in \ref{appendix_architecture}).\\ 
The dataset is an extension of the dataset used in \cite{Accou2021ModelingTR, Accou2021PredictingSI} with 90 subjects rather than 48 before.
In this paper a variation on the match-mismatch paradigm explained above is used: the model takes one EEG segment and one speech segment (the speech envelope) as inputs and has to decide whether they match or not. 
On the speech stream, the dimensionality reduction block consists of a convolutional layer followed by an LSTM layer. As the LSTM memory capacity is limited, the CNN is used as a preprocessing layer that reduces the number of recurrence steps (kernel of more than 1 sample slides along the time axis and reduces its dimension) the LSTM will have to perform. 
On the EEG stream, the dimensionality reduction block consists of a convolutional layer, followed by two dense layers.
Both EEG and the speech segment are hence projected into a common embedded space.
Similarly to the dilated-convolutional model from \cite{Accou2021ModelingTR}, the cosine similarity is here used as the correspondence measure, except that this time it is computed along the time axis (i.e., the output will be a vector with the length being the number of time samples).
The decision layer block is composed of a time-distributed dense layer followed by a calculation of the mean to select the outcome (i.e., matched or mismatched). The data split for training, validation, and test is 80:10:10 as in \cite{Accou2021ModelingTR}.\\
As the dilated-convolutional model presented in the two previous papers, this LSTM-based architecture outperforms a CNN baseline as well as a linear decoder baseline (85\%, 73\%, and 69\% classification accuracy respectively). Subject-independent training leads to higher accuracy than subject-dependent with this model which is a great advantage over linear models and avoids re-training to evaluate the performance of new subjects.\\
The same architecture was used with other speech features in \cite{Monesi2021INTERSPEECH}, such as such as the Mel spectrogram, voice activity, phoneme identity, and word embedding.
The results suggest that the model exploits information about silences, intensity, and broad phonetic classes from the EEG. Furthermore, the Mel spectrogram, which contains all this information, yields the highest accuracy (84\%) among all the features.\\
As opposed to \cite{Accou2021ModelingTR, Accou2021PredictingSI}, no negative sample is taken for the MM task. A negative sample is added in \cite{Monesi2021INTERSPEECH}, but in both studies, it remains a binary classification that raises the same criticism mentioned for \cite{Accou2021ModelingTR, Accou2021PredictingSI}.\\


\noindent\textit{Deep Canonical Correlation Analysis For Decoding The Auditory Brain (Katthi et al., 2020):}
Canonical correlation analysis (CCA) is a linear method to project two signals to a latent space that maximizes the correlation between the two signals. 
Following-up on \cite{DECHEVEIGNE2018}, \cite{Katthi2021DeepMC} introduce a deep CCA (DCCA) model for speech-EEG data. This model has shown improvement over linear CCA (LCCA) on image data and is expected, along with appropriate regularization methods, to improve the resulting correlations on speech-EEG data.\\
The dataset used is from \cite{DiLiberto2015}. 6 subjects listened to single-speaker audio books within 20 trials of duration 160~s). 19 trials are used for training and 1 for testing. Within the training set, 90\% is allocated for training and 10\% for validation.
The authors tried two neural network architectures for the deep CCA models. The first architecture contains a 2 hidden layer network for each of the envelope and the EEG side, followed by a 1-dimensional output layer. The second architecture contains 4 hidden layers. They use a leaky ReLU activation function, with a negative slope coefficient of 0.1 at the output of the deep CCA model.
The neural network is trained to maximize the correlation between the embedded representations of speech and EEG.
The first architecture consistently outperformed the second one so only these results will be mentioned here. The authors experimented with several configurations of the deep CCA model and show that the model consistently outperforms the linear counterpart (best subject correlation: 0.4 and 0.31; worst subject correlation: 0.21 and 0.18 for DCCA and LCCA respectively).\\
The comparison between the linear CCA's performance and the deep CCA's performance is evaluated in a subject-specific paradigm. Hence, the model is not trained in the perspective to generalize across subjects, which is one of the expected improvements from deep learning models over linear models (see Section \ref{Pitfalls}).
Another issue is the difficulty to interpret CCA correlation coefficients: the correlation between the EEG and speech latent representations is hardly comparable to the correlation between the predicted and ground truth signals usually reported in reconstruction/prediction studies.\\

\noindent\textit{Deep Correlation Analysis for Audio-EEG Decoding (Katthi et al 2021):}
The authors develop two models: one for intra-subject audio-EEG correlation analysis (DCCA model introduced in \cite{Katthi2021DeepMC}) and another for an inter-subject configuration (the deep multiway canonical correlation analysis, or DMCCA).
Intra- and inter-subject correlation analyses attempt to suppress the EEG artifacts while preserving the components related to the stimulus. DCCA uses the intra-subject information to do so, while DMCCA uses information shared across subjects to attempt EEG denoising.\\
The authors here compare 4 architectures and evaluate the Pearson correlation between the transformed EEG and audio signals. Two possible inter-subject architectures are firstly used: linear-multiway canonical correlation analysis (LMCCA) and DMCCA; then two possible intra-subject architectures: LCCA and DCCA. This leads to 4 possible combinations: LMCCA/LCCA, LMCCA/DCCA, DMCCA/LCCA, and DMCCA/DCCA) from which performances are compared.\\
The dataset used is from \cite{DiLiberto2015}. 8 subjects listened to single-speaker audiobooks within 20 trials of duration 160~s). \\
The structure of DCCA is explained in \cite{Katthi2021DeepMC}. LCCA and LMCCA \citep{DECHEVEIGNE2019} are used here as linear baselines.
The DMCCA model is an autoencoder finding a latent representation of each EEG with $d$ dimensions ($d$ is a hyperparameter to optimize) and reconstructing the original responses from them. The encoder has two hidden layers and an output layer of $d$ units. The decoding part has two hidden layers.\\
For a majority of the subjects, the authors found that the deep learning combination DMCCA/DCCA improved the Pearson correlation significantly over the other linear combination LMCCA/LCCA (average correlation for DMCCA/DCCA: 0.344, LMCCA/LCCA: 0.270). As a conclusion, more optimal transforms can therefore be found using deep models.\\
We address one drawback of this method: the deep CCA model projects the EEG and speech signals into a common latent space that maximizes the correlation between them. Such correlations are higher for low-frequency bands \citep{Etard2019}, which might bias the model to omit some higher-frequency band information. It becomes thus challenging to compare correlations obtained with CCA and regression tasks correlations computed between the ground truth and the reconstructed/predicted signals.\\

\noindent \textit{Robust decoding of the speech envelope from EEG recordings through deep neural networks (Thornton et al. 2022):}
In \cite{thornton2022robust}, the authors propose two neural network structures to reconstruct the envelope from EEG recordings in different listening conditions.\\ 
Data is included from 13 subjects who listen to a single-speaker audiobook, in noiseless and anechoic conditions, for a total of 40 minutes per subject, recorded in 15 different trials. 9 of these trials were used for training, 3 for evaluation, and 3 for testing.\\
As a baseline, the authors use a linear backward model. Two neural networks are compared. The first is an FCNN. A segment of 50-time samples (400~ms) is first flattened and then put through $L$ (a hyperparameter to optimize) fully connected layers, with each layer containing fewer neurons than the preceding layer. The output of the final layer is a scalar, representing an estimate of the envelope at the onset of the segment. 
The second network is a CNN inspired by EEGnet \citep{Lawhern_2018} and performs a temporal convolution on the input, followed by a spatial convolution and a depthwise separable convolution. The output is then flattened and reduced to a single scalar output by taking a linear combination.\\
For both the subject-independent and subject-specific training paradigms, both the FCNN and the CNN models significantly outperformed the linear baseline (subject-independent: median Pearson $r=0.12$, $r=0.13$ and $r=0.08$ for the FCNN, CNN and linear baseline respectively; subject-specific: median Pearson $r=0.22$, $r=0.22$ and $r=0.18$ for the FCNN, CNN and linear baseline respectively).\\
As reported in \cite{Accou2023}, correlation distribution of the FCNN and CNN models remained very similar when evaluated on unseen stimulus segments from the same subjects used during training (median: $r=0.14$ and $r=0.14$ respectively), but also when evaluating it on different subjects from the DTU dataset (median: $r=0.11$ and $r=0.14$). These findings indicate a good generalization of the model across subjects and speech content (see Section \ref{SI_model}).\\


The LSTM-based model, for both spectrogram and envelope, led to higher Pearson correlation values than the linear baseline. For native speakers, the mean correlation across subjects increased from 0.15 to 0.20 for the envelope, and from 0.07 to 0.09 for Mel spectrogram. 

As far as we can infer from the article, the LSTM-based model was trained in a subject-specific paradigm. Considering the memory capacity of LSTM layers, the model could simply remember characteristics from a subject such as EEG electrodes placement. A solution is to train the model on EEG responses from multiple subjects listening to a common stimulus to compensate for these differences (see Section \ref{SI_model}).\\

The three next papers \citep{Krishna2020, KrishnaEUSIPCO2020, Krishna2021NER} are from the same research group, and they use the same dataset, we therefore discuss them together. \\
The dataset consists of 4 subjects listening to 4 different natural utterances (2-word sentences). The authors collected 70 speech-EEG recordings per subject per sentence. As far as we can infer one recording corresponds to the EEG response to one sentence. The train:validation:test ratio was 80:10:10 and supposedly constant across subjects.\\ 

\noindent \textit{Speech synthesis using EEG (Krishna et al. 2020):}
Reconstructing (e.g., imagined) speech from EEG is a way to envision a brain-computer interface (BCI) to enable communication when overt speech is impossible.  Speech synthesis was attempted in \cite{Krishna2020}. This paper aims at developing deep learning architectures with recurrent layers such as GRUs to reconstruct presented continuous speech or listening utterances from their EEG.\\
The speech synthesis model first has two GRU layers. The last GRU layer is connected to a time-distributed dense layer with 13 units which corresponds to the number of speech features to predict at every time step (i.e., Mel frequency cepstrum coefficient, or MFCC). The model was evaluated using two metrics: the Mel cepstral distortion (MCD) and the root-mean-squared error (RMSE) between the predicted and ground truth MFCC. \\
On both continuous speech and listening utterances, the presented GRU-based speech synthesis model outperformed LSTM-based models introduced in \cite{Krishna2020ARXIV} (which was not published). We do not report the performance figures here as they are absolute RMSE and MCD values which makes it complicated to compare with other papers.\\

\noindent \textit{Generating EEG features from Acoustic features (Krishna et al. 2020, EUSIPCO):}
In \cite{KrishnaEUSIPCO2020}, the authors developed an RNN-based forward regression model. It can be seen as the inverse problem of the EEG-based synthesis from \cite{Krishna2020}. They present two models, a regression and a generative adversarial network (GAN) to predict EEG from acoustic features (MFCC).\\
The regression model has a similar structure as in \cite{Krishna2020} for speech synthesis. MFCC is fed into two GRU layers, followed by a time-distributed dense layer with the number of units corresponding to the dimensions of the EEG feature set used. An alternative regression model with bi-GRU layers instead of GRU layers was also evaluated.\\
A GAN \citep{NIPS2014_GAN} architecture was also built to predict EEG from acoustic features. The motivation to use a GAN is that the loss function is learned whereas in a regression model the loss is fixed (here MSE).
The generator is here made of two bi-GRU layers followed by a time-distributed dense layer with the number of units corresponding to the number of EEG feature set dimensions. The output of the generator is referred to as fake EEG.
The discriminator applies two bi-GRU layers connected in parallel. At each training step, a pair of inputs is fed into the discriminator. The discriminator takes pairs of (real MFCC features, fake EEG) and (real MFCC features, real EEG). The outputs of the two parallel Bi-GRUs are concatenated and then fed to a GRU layer. The last time step output of the GRU layer is fed into the dense layer with a sigmoid activation function.
If appropriately trained, the generator should be able to generate realistic EEG corresponding to a given MFCC input, thus fulfilling the same task as a forward model.\\
The authors obtained a lower RMSE than in the speech synthesis task \citep{Krishna2020} and therefore demonstrated that using their dataset with their processing steps, it was easier for a recurrent deep learning model to predict EEG from acoustic features than the inverse task.\\

\noindent \textit{Advancing Speech Synthesis using EEG (Krishna et al. 2021):} 
In \cite{Krishna2021NER}, the authors attempt speech synthesis using an attention-regression model (i.e., a regression model with an attention mechanism, AR). Their contribution consists of many points: first, they use an AR model to predict acoustic features from EEG features. 
Secondly, they use a two-step approach to solve the same task (i.e., one AR model to predict articulatory features from EEG features, then another AR to predict acoustic from articulatory features).
Thirdly, they propose a deep learning model that takes raw EEG waveform signals as input and directly produces audio waveform as output.
Finally, they demonstrate predicting 16 acoustic features from EEG features.\\
The architecture of the AR model is as follows: first, an encoder GRU layer, then a Luong dot product attention layer \citep{luong-etal-2015-effective}. The context vectors obtained are then provided to the decoder GRU layer. The decoder GRU outputs are passed to a time-distributed dense layer with a linear activation function.\\
The architecture for speech synthesis of an audio waveform using EEG is as follows: the input EEG is fed into a temporal convolutional network (TCN) layer. Features extracted by the TCN are up-sampled. A TCN layer is then applied before the final up-sampling layer. The up-sampled features are then passed to a time-distributed dense layer consisting of a linear activation function which directly outputs the audio waveform.\\
The last model to predict 16 acoustic features from EEG features was developed. It is a GRU-based regression model. First, the EEG input is provided to a GRU layer, then passed through a time-distributed layer with a linear activation function.\\
The results presented in this paper show how different acoustic features are related to EEG recorded during speech perception and production. Their newly introduced AR models outperform the regression model they introduced in \cite{Krishna2020} (i.e., higher MCD for the new model on a majority of subjects). They also show some RMSE values obtained per subject in the raw EEG to audio waveform task but without comparison.\\

We address a common issue for \cite{Krishna2020, KrishnaEUSIPCO2020, Krishna2021NER}: the stimulus content is limited. It is tricky to evaluate the backward/forward modeling performance of a model with four utterances. First, it might be that the model overfits completely on these utterances as they are very short (2 words per utterance). In addition, considering the 80:10:10 ratio used for each subjects, the model certainly saw all the utterances during training, which comforts the overfitting hypothesis.\\
In \cite{KrishnaEUSIPCO2020}, the authors obtained better performance for EEG prediction than for speech reconstruction. This finding is surprising as it is theoretically possible to reconstruct the MFCC features perfectly from the EEG, however considering that the EEG contains speech-unrelated components, it is normally impossible to predict the whole EEG signal from speech.\\

\noindent\textit{Learning subject-invariant representations from speech-evoked EEG using variational autoencoders (Bollens et al. 2021):}
The authors introduce a model aiming to disentangle subject-invariant and subject-specific information from speech-evoked EEG using variational autoencoders (VAEs). For that purpose, EEG is modeled with two disentangled latent spaces (i.e., one that models subject information and one content information). The disentanglement accuracy is then measured using subject classification, and content classification tasks from the latent representations learned by the VAE. 
\\
The dataset is an extension of the one used in \cite{Accou2021ModelingTR}, and consists of 100 normal-hearing native Flemish subjects. Each listened to a minimum of 6 up to 8 stories (each around 14 minutes).
The presented model is a factorized hierarchical variational autoencoder (FHVAE) encoding EEG signals in two disentangled latent spaces. One captures high-level slow-varying information, while the other captures residual fast-changing information. It aims at modeling neural responses for short EEG segments (i.e., 500~ms). $z_{1}=\mu_{1} + \epsilon\sigma_{1}$ is defined as the latent content variable (i.e., containing content-information) whereas $z_{2}=\mu_2 + \epsilon\sigma_{2}$ is defined as the latent subject variable (i.e., containing subject information). $\mu_{k}$ is the conditional mean and $\sigma_{k}$ the conditional variance, with k=1,2, the index of the corresponding latent variable.\\
The architecture is as follows: first, in the encoder, $z_1$ and $z_2$ are predicted by a stacked LSTM network of two layers, followed by two separate single fully connected layers, predicting the conditional mean $\mu_{k}$ and variance $\sigma_{k}$, respectively. 
The second part is a decoder, a two-layer stacked LSTM decoder network, which is fed at each time step with the concatenation of the sampled $z_{1}$ and $z_{2}$ from the posterior distribution. Subsequently, two separate single fully-connected layers take the output from the LSTM layers and predict the probability distribution of the corresponding time frame.
For more details about the implementation, the authors refer to \cite{NIPS2017_3f5ee243}. The model is trained first to learn subject information (i.e., FHVAE), and then $z_{1}$ regularization to enhance content separation is added to a second model (i.e., extended FHVAE). The dataset split for training, validation and test sets is 80:10:10 respectively as used in \cite{Monesi2020, Accou2021ModelingTR}.\\
Subject-classification accuracy reached 98.94\% and 98.96\% for the latent $z_{2}$ representations of respectively the FHVAE and the Extended FHVAE architecture, suggesting both models succeeded at extracting subject information. On the other hand, subject-classification accuracy for the latent $z_{1}$ representation was around 2\%, which suggests very good disentanglement.\\
The extended FHVAE improves content classification from $z_1$ from 53.89\% to 62.91\% on a binary classification task (chance level $50\%$), while the classification performance from $z_2$ decreases, confirming that the extended model succeeds at modeling content-generating factors in $z_{1}$ but not in $z_{2}$. 
These results are a first step towards making deep learning models relating speech to EEG either generalizable, which, considering the idiosyncratic nature of EEG across subjects, is highly relevant and necessary.\\

\noindent\textit{Sequential Attention-based Detection of Semantic Incongruities from EEG While Listening to Speech (Motomura et al. 2020):}
A classification paradigm is used to detect semantic incongruities in sentences from EEG recordings. In this paper, the authors use an attention-based recurrent neural network (ARNN) to detect whether a given sentence contains an anomalous word or not.\\
In this study, 19 subjects listened to 200 sentences in Japanese: 40 semantically correct, 40 semantically incorrect, 40 syntactically correct, 40 syntactically incorrect, and 40 filler sentences. The concatenated data from 13 subjects were used for training, from 2 subjects for the validation set, and 4 subjects for the test set.\\
The authors evaluated two models: a bi-GRU layer with or without an attention mechanism respectively. The attention mechanism provides attention weights to each time point with the hypothesis that they might not all have the same importance for the classification (e.g., around the onset of the anomalous word in the sentence, the attention weights have a higher magnitude). The output was obtained using a weighted sum over all time points with the attention weights.\\
The accuracy obtained on the binary classification reaches 63.5\% which is statistically above the chance level, as well as previous models including anomalous word onset (50.9\%).\\

\noindent\textit{Keyword-spotting and speech onset detection in EEG-based Brain-Computer Interfaces (Sakthi et al. 2021):}
Three tasks were investigated using an LSTM- and GRU-based network: a sentence spotter (SS), a phoneme vs silence classification (PS), and an audio vs audio-visual stimuli classification (AV). The performance of a deep neural network with recurrent layers is investigated to further be integrated into BCI systems.\\
The dataset consists of 16 native English speakers, they listened to four blocks of 125 sentences (all unique sentences) and the fifth block of 100 sentences (10 unique sentences repeated). The same 70\% of the stimulus data per subject was used for training and the remaining 30\% for testing.\\
The SS model is a binary classification model to predict whether the input is the start of a sentence or not. 
The EEG signal is segmented (from 500ms before to 250ms after the sentence onset) and a PCA is applied.
The resulting embedded representation is then fed to two subsequent GRU layers. Finally, a dense layer is applied with one hidden unit and a sigmoid function. Alternatively, a second version of the model was trained, with LSTM layers instead of the GRU layers.\\
The PS model is a binary classification model to predict whether the input is a phoneme or a silence. 
A PCA is applied to the segmented EEG input The resulting embedded representation is then fed to two GRU layers and afterward a dense layer was applied with a sigmoid as an activation function. 
A similar model was trained using LSTM instead of GRU layers.\\
The AV model is a binary classification model to identify if a given EEG response was evoked by audio or audio-visual stimuli. The EEG signal is segmented in 3~s chunks and a PCA is applied as for the previous models. The model has 4 GRU layers. The resulting output was processed by a fully connected layer of size 1 with sigmoid activation.
As a baseline, a Naive Bayes model was trained on all the tasks listed above.\\
The recurrent architectures consistently outperformed the simple naive Bayes model for all tasks. All three models gave a better performance with the GRU layer rather than the LSTM layers. We here do not provide the exact figures as the SS and PS models are evaluated using the F1 score which cannot be compared to the other studies reviewed here. The reason why the accuracy values are so high is because of the rare presence of onsets in data, hence the use of F1 scores. The AV classifier obtained an accuracy of 98.15\%.\\

\end{document}